\documentclass[acmsmall,screen]{acmart}
\AtBeginDocument{%
  }

\usepackage{mathrsfs}
\usepackage{subfigure}
\usepackage{microtype,flushend}
\usepackage{colortbl}
\usepackage{balance}
\usepackage[most]{tcolorbox}
\usepackage{multirow}
\usepackage{fontawesome}
\usepackage{threeparttable}
\usepackage{color}
\usepackage[ruled,vlined]{algorithm2e}
\usepackage{xspace}
 \setcopyright{none}
\renewcommand\footnotetextcopyrightpermission[1]{}

\newcommand{\tabincell}[2]{\begin{tabular}{@{}#1@{}}#2\end{tabular}}

\usepackage[framemethod=tikz]{mdframed}
\newcounter{finding}
\newcommand{\finding}[1]{\refstepcounter{finding}
  \vspace{2.3mm}
 \begin{mdframed}[linecolor=gray,roundcorner=12pt,backgroundcolor=gray!15,linewidth=3pt,innerleftmargin=2pt, leftmargin=0cm,rightmargin=0cm,topline=false,bottomline=false,rightline = false]
  \textbf{Ans. to RQ\arabic{finding}:} #1
 \end{mdframed}
 \vspace{2.3mm}
}

\lstdefinestyle{mystyle}{
    numberstyle=\tiny,
    basicstyle=\ttfamily\footnotesize,
    breakatwhitespace=false,         
    breaklines=true,                 
    captionpos=b,                    
    keepspaces=true,                 
    numbers=left,                    
    numbersep=5pt,                  
    showspaces=false,                
    showstringspaces=false,
    showtabs=false,                  
    tabsize=2,
    frame={bottomline}
}
\setcopyright{acmcopyright}
\copyrightyear{2025}
\acmYear{2025}
\acmDOI{XXXXXXX.XXXXXXX}

\begin{document}

\newcommand{\toolName}{\textit{SCOPE}\xspace}
\newcommand{\compareOne}{\textit{PT4Cloud}\xspace}
\newcommand{\compareTwo}{\textit{Metior}\xspace}
\newcommand{\compareThree}{\textit{CONFIRM}\xspace}

\title{\toolName: Performance Testing for Serverless Computing}


\author{Jinfeng Wen}
\affiliation{
  \institution{Beijing University of Posts and Telecommunications}
  \city{Beijing}
  \country{China}}
\email{jinfeng.wen@bupt.edu.cn}

\author{Zhenpeng Chen}
\affiliation{
  \institution{Nanyang Technological University}
  \city{Singapore}
  \country{Singapore}}
\email{zhenpeng.chen@ntu.edu.sg}

\author{Jianshu Zhao}
\affiliation{
  \institution{Beijing University of Posts and Telecommunications}
  \city{Beijing}
  \country{China}}
\email{jianshuzhao@bupt.edu.cn}

\author{Federica Sarro}
\affiliation{
  \institution{University College London}
  \city{London}
  \country{United Kingdom}
}
\email{f.sarro@ucl.ac.uk}

\author{Haodi Ping}
\affiliation{
  \institution{Beijing University of Technology}
  \city{Beijing}
  \country{China}}
\email{haodi.ping@bjut.edu.cn}

\author{Ying Zhang}
\affiliation{
  \institution{Peking University}
  \city{Beijing}
  \country{China}}
\email{ying.zhang@pku.edu.cn}

\author{Shangguang Wang}
\affiliation{
  \institution{Beijing University of Posts and Telecommunications}
  \city{Beijing}
  \country{China}}
\email{sgwang@bupt.edu.cn}

\author{Xuanzhe Liu}\authornote{Corresponding author}
\affiliation{
  \institution{Peking University}
  \city{Beijing}
  \country{China}}
\email{liuxuanzhe@pku.edu.cn}







\renewcommand{\shortauthors}{Wen et al.}

\begin{abstract}
Serverless computing is a popular cloud computing paradigm that has found widespread adoption across various online workloads. It allows software engineers to develop cloud applications as a set of functions (called \textit{serverless functions}). However, accurately measuring the performance (i.e., end-to-end response latency) of serverless functions is challenging due to the highly dynamic nature of the environment in which they run. To tackle this problem, a potential solution is to apply checks of performance testing techniques to determine how many repetitions of a given serverless function across a range of inputs are needed to cater to the performance fluctuation. However, the available literature lacks performance testing approaches designed explicitly for serverless computing. In this paper, we propose \toolName, the first \underline{s}erverless \underline{c}omputing-\underline{o}riented \underline{p}erformance t\underline{e}sting approach.
\toolName takes into account the unique performance characteristics of serverless functions, such as their short execution durations and on-demand triggering. As such, \toolName is designed as a fine-grained analysis approach. 
\toolName incorporates the accuracy check and the consistency check to obtain the accurate and reliable performance of serverless functions. The evaluation shows that \toolName provides testing results with 97.25\% accuracy, 33.83 percentage points higher than the best currently available technique. Moreover, the superiority of \toolName over the state-of-the-art holds on all functions that we study.
\end{abstract}





\maketitle

\section{Introduction}\label{sec:introduction}
Serverless computing is becoming a mainstream cloud computing paradigm that has been widely adopted in various online workloads like big data analytics, deep learning, large language models (LLM), and so on~\cite{fouladi2017encoding-18,wu2022Serverless-21,muller2020lambada-23, wu2023qos, liu2024faasgraph, wu2024loongserve, wu2024dlora, zhong2024distserve}. It frees software engineers from tedious and error-prone infrastructure management and allows them to focus on developing a cloud application as a set of event-driven functions, called \textit{serverless functions}~\cite{Wen21challenges}. To support the execution of serverless functions, mainstream cloud vendors have provided serverless platforms, such as AWS Lambda~\cite{aws} and Google Cloud Functions~\cite{google}. It is predicted that the market size of serverless computing is projected to grow significantly, reaching $\$$41 billion in 2028, compared to $\$$19 billion in 2024~\cite{marketreport}.
It indicates that an increasing number of developers will pivot to develop serverless functions.

The surging popularity of serverless computing has led to heightened interest among diverse research communities, including Software Engineering (SE) and Systems~\cite{wen2022literature}. Notably, performance stands out as the most extensively studied aspect in serverless computing research~\cite{wen2022literature}. However, it is challenging to obtain accurate and reliable performance (i.e., end-to-end response latency) measurements for serverless functions due to the following reasons. Serverless platforms, where serverless functions are executed, have a highly dynamic cloud underlying infrastructure. This introduces various challenges for accurate and reliable performance measurement such as multi-tenancy, resource management, and networking issues~\cite{mahgoub2022orion-36, fuerst2022locality-32, perron2020starling-26, wu2022container, huang2023duo, liu2024muxflow, gu2023elasticflow}. Serverless functions are generally short-lived tasks that require a small memory size to be configured to provide the resource~\cite{ShahradATC20, singhvi2021atoll}. This results in a high-density deployment environment, increasing the risk of performance fluctuations~\cite{zhao2021understanding-34, patterson2022hivemind-11, wen2025unveiling}.
Under these circumstances, serverless functions can produce highly fluctuating performance results even with multiple identical runs when executed on serverless platforms. 
Moreover, developers design a variety of serverless functions with different functionalities, possibly with different levels of performance fluctuations.

To alleviate this issue, a straightforward way is to set a fixed number of measurement repetitions for all evaluated serverless functions, according to experiment repetitions used in prior studies. 
The results obtained from measurements are used as the accurate and reliable performance for the serverless function.
However, due to the diversity of serverless functions developed by various developers, it is evident that not all serverless functions require or benefit from the same level of repetition.
Therefore, using a fixed number of repetitions is unreasonable and ineffective in determining the actual performance for different serverless functions executed on different platforms.
A better strategy would be to devise a method for recommending a customized number of repetitions for each serverless function, thereby achieving a more accurate and reliable performance evaluation.

To achieve this goal, a possible solution is to use performance testing techniques, standard procedures for obtaining and evaluating the performance of a software application in SE~\cite{zhao2019localized, burger2019bottleneck, he2021performance}. Generally, the performance testing technique is conducted by repeatedly executing the application-under-test with a set of inputs until a stopping criterion deems that the performance results obtained from the test are accurate~\cite{mostafa2017perfranker, maricq2018taming, uta2020big, he2019statistics, he2021performance, alghmadi2016automated}. 
Performance testing techniques are important for serverless computing. Serverless functions can be part of user-facing features where end-to-end response time directly impacts user experience, thus demanding performance-critical considerations. Other serverless functions can be invoked infrequently, such as end-of-month reports or daily reminder emails. Although such functions are often tolerant of cold start latencies, characterizing their performance remains crucial to ensure that service-level objectives (SLOs) are met and to make informed cost-benefit decisions regarding their deployment. Given the diversity of use cases, it is necessary for the developers of serverless functions to employ effective performance testing techniques.
However, to the best of our knowledge, the literature lacks a performance testing approach tailored to serverless functions.

In this paper, we propose \toolName, the first \underline{s}erverless \underline{c}omputing-\underline{o}riented \underline{p}erformance t\underline{e}sting approach.
\toolName takes into account the unique performance characteristics of serverless functions, such as their short execution durations and on-demand triggering.
Our primary goal with \toolName is to provide a novel stopping criterion to determine a specific repetition number to obtain highly accurate and reliable performance profiles for each given serverless function.
\toolName poses a strict requirement on accuracy and utilizes the accuracy check and the consistency check to determine the stopping criterion for repeated runs. 
The accuracy check first utilizes the non-parametric confidence interval to analyze whether a specific performance profile is accurate. For the performance result set of the current test, if most of its performance is determined to be accurate, this set is considered accurate.
The consistency check examines whether the performance result set of the current test and the performance result set obtained from the previous run intervals are both accurate. If this is true, \toolName deems that the performance result set of the current test is sufficient to represent the accurate and reliable performance of the serverless function and terminate the repeated runs.

We evaluate the effectiveness of \toolName and state-of-the-art techniques developed for traditional cloud applications by investigating 65 serverless functions from existing work~\cite{wen2023revisiting}. We use the performance results of 1,000 identical runs (which is the largest number of runs used in the literature) of a serverless function as its ground-truth performance.
The evaluation of the 65 serverless functions shows that \toolName provides testing results with 97.25\% accuracy, 33.83 percentage points higher than the best currently available technique. Furthermore, \toolName is widely effective, as it outperforms state-of-the-art techniques on all serverless functions that we consider.
In contrast to the indiscriminate implementation of a fixed repetition strategy, \toolName shows enhanced flexibility and efficacy in determining a specific repetition and achieving accurate and reliable performance across diverse serverless functions.

To the best of our knowledge, our research is the first to explore both (1) a performance testing technique specifically tailored for serverless computing and (2) an empirical study on the effectiveness of performance testing techniques in serverless computing. These contributions constitute the \textit{novelty} of our work.
We also provide a public repository~\cite{ourdata} including all the data and code used in this study to facilitate future replication and extension.

\section{Background}\label{sec:background}


\subsection{Serverless computing}






\begin{figure}[t]
	\centering
    \includegraphics[width=0.5\textwidth]{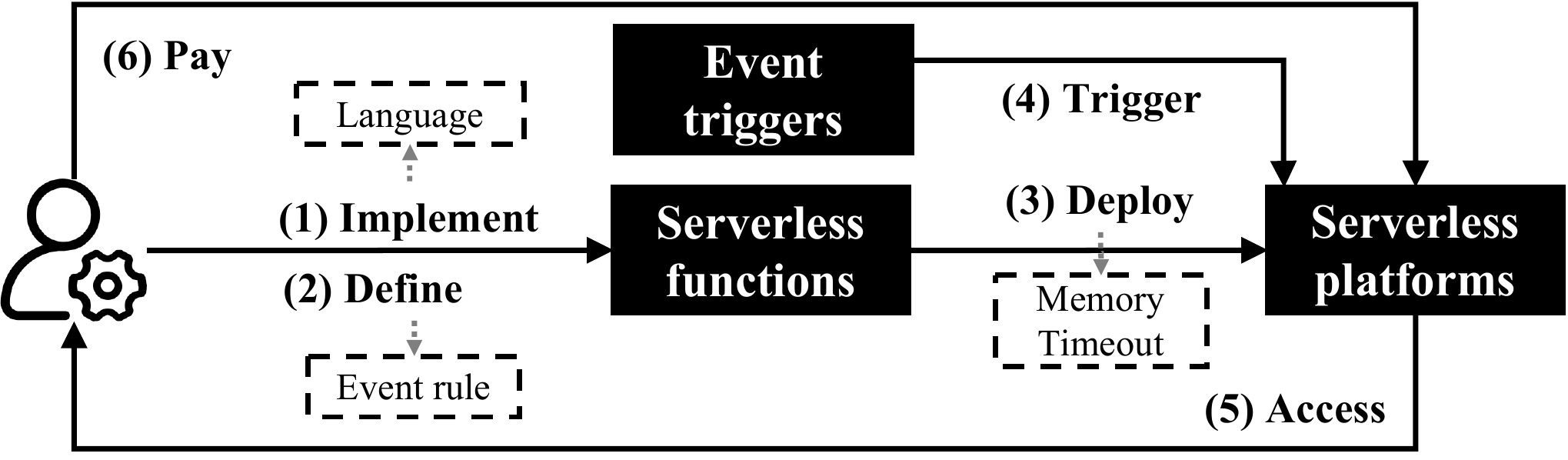}
    \caption{The process of using serverless computing.}
    \label{fig:DevelopmentslsFun}
\end{figure}


In serverless computing, developers focus on application implementation based on serverless functions.
Fig.~\ref{fig:DevelopmentslsFun} depicts the process of using serverless computing for developers. 
(1) First, developers implement event-driven serverless functions using high-level programming languages, e.g., Python and JavaScript~\cite{serverlesscommunitysurvey, serverlesscommunitysurvey1, eskandani2021wonderless, eismann2021state}.
(2) Second, developers can define specific rules that bind their functions to the corresponding events, e.g., HTTP requests and data updates in cloud storage. 
(3) Then, serverless functions are deployed to the serverless platform along with dependent libraries, e.g., \textit{Numpy}. During this phase, developers can provide specific function configurations, e.g., memory size and timeout time~\cite{wen2022literature, hassan2021survey}. 
(4) When the serverless function is triggered by predefined events, the serverless platform automatically launches new function instances (e.g., containers or virtual machines) or reuses existing ones to process requests. 
(5) Upon completion of executions, the serverless platform logs information related to function execution, allowing developers to access and review it later.
(6) Finally, developers pay for the cost according to the number of requests and the resources actually allocated or consumed by the serverless function~\cite{lin2020modeling, wen2022literature}.



\subsection{Serverless function performance}

The performance of serverless computing has gained widespread attention in the serverless computing literature~\cite{wen2022literature, li2022serverless, mampage2021holistic, hassan2021survey}. Researchers have proposed various solutions to optimize serverless function performance~\cite{lin2020modeling, singhvi2021atoll, ShahradATC20, wen2023FaaSlight, eismann2021sizeless-183, xu2019adaptive-100, wu2022container-123}. 
Serverless function performance can be classified into two types: cold-start performance and warm-start performance. 
When the serverless function is executed on newly launched instances in the serverless platform, it will produce cold-start performance.
If the serverless platform has reusable instances for the same function to handle requests within a short keep-alive time (e.g., 7 minutes for AWS Lambda~\cite{awsIdlelifetime}), the serverless function will produce warm-start performance.

The serverless function performance that we focus on is end-to-end response latency, i.e., the time period between sending a request to invoke a serverless function hosted in the serverless platform and receiving the execution result of the function. End-to-end response latency is the most common metric used for performance evaluation and optimization of serverless functions~\cite{mahgoub2022orion-36, zhao2021understanding-34, eismann2021state, wen2021characterizing, Wen21challenges}. It contains the serverless platform's preparation time, the function's task processing time, etc. In this paper, we study both cold-start and warm-start end-to-end response latencies.

\subsection{Motivation}

We present examples of serverless function performance across multiple runs. For instance, we observe \textit{Func52} and \textit{Func30}, which are part of our dataset described in Section~\ref{sec:dataset}. The maximum and minimum values of the cold-start end-to-end response latencies of \textit{Func52} can vary by as much as 45.83\% between runs. Similarly, the warm-start end-to-end response latencies of \textit{Func30} can differ by  369.24\%, with the maximum value being 4.69 times greater than the minimum. These examples show the significant variability in serverless function performance, even with repeated, identical runs.
Furthermore, as shown in Fig.~\ref{fig:compareFixed}, the bars represent the number of repetitions needed to achieve reliable performance across different serverless functions. The results demonstrate that each serverless function requires a tailored number of repetitions to ensure accuracy and reliability.  Overall, these observations emphasize the need for a method that recommends a customized number of repetitions for each serverless function to obtain accurate and reliable performance.


\section{Our Performance Testing Approach: \toolName}\label{sec:approach}

It is crucial to obtain accurate and reliable performance for serverless functions. However, the existing serverless computing literature does not offer performance testing techniques specifically for it.
Therefore, designing a performance testing approach for serverless computing is necessary. Generally, the primary goal of a performance testing approach is to determine whether the tested performance results accurately reflect the actual performance, guiding the decision on whether additional repeated runs from another run interval are necessary.
The run interval refers to the specific number of repetitions or trials needed for executing the serverless function. Within each run interval, the serverless function is executed multiple times to acquire new performance data samples.
To motivate our approach design, we first summarize key characteristics of serverless functions.

\subsection{Key characteristics}

\noindent  $\bullet$ \textit{Serverless functions run for short duration.} 
The response latency of serverless functions is often measured in milliseconds~\cite{ShahradATC20, zhao2021understanding-34}. 
However, short-lived serverless functions executed on serverless platforms with highly dynamic cloud underlying infrastructure can exhibit significant performance fluctuations~\cite{wen2023revisiting, zhao2021understanding-34, patterson2022hivemind-11}.





\noindent $\bullet$ \textit{Serverless functions are executed at small run intervals.} 
In one run interval of performance testing, which refers to the specific number of repetitions required to execute the serverless function, the number of performance results obtained is generally small.
This is because the serverless computing scenario inherently uses small repetitions. 
(1) Serverless functions can be triggered at any time, allowing developers and researchers to obtain performance results as needed. Small repetitions serve as the foundation, as developers and researchers pay based on the number of invocations and resource consumption. Consequently, they tend to invoke serverless functions with small repetitions on demand.
(2) In research work on serverless computing, prior experimental evaluations have employed a predefined number of runs performed on serverless functions for performance analysis. This number is generally small repetitions~\cite{wen2025unveiling}, e.g., 3, 10, and 20 times, thus obtaining a small number of performance results.


Based on these key characteristics, there is a pressing need for a fine-grained, high-accuracy performance testing approach specifically tailored to serverless computing. Such an approach would facilitate accurate and reliable performance measurements and obtain a specified number of repetitions for serverless functions.

\begin{figure*}[t]
	\centering
    \includegraphics[width=0.95\textwidth]{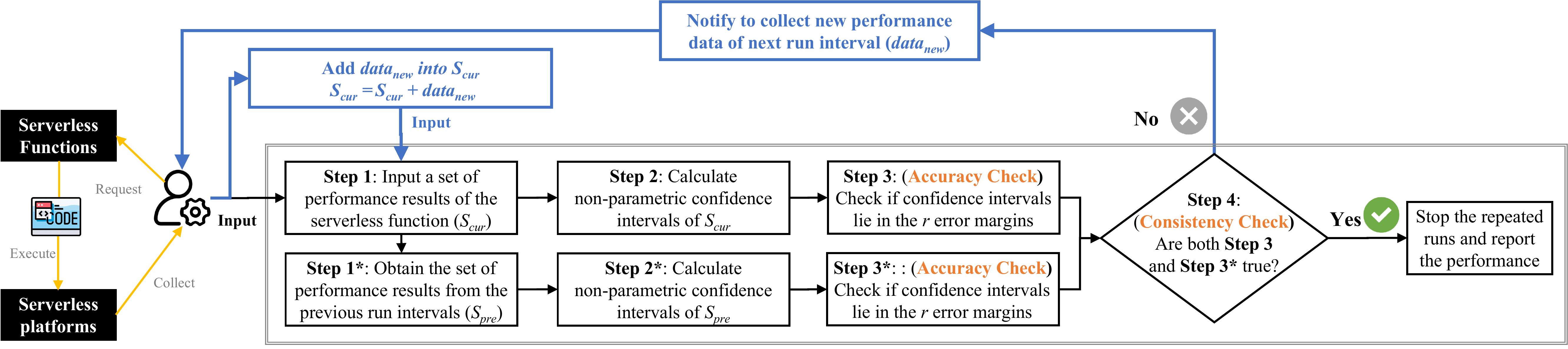}
    \caption{The workflow of \textbf{\toolName}.}
    \label{fig:overview}
\end{figure*}


\subsection{Overview of \toolName}






We propose \toolName, a performance testing approach for serverless computing. 
\toolName is an automated performance tester that provides accurate and reliable performance for serverless functions.
To use \toolName, developers collect the performance result set of a function with a given input from the serverless platform and provide it to \toolName. \toolName determines whether this set accurately reflects real performance and whether more repeated runs from another run interval are necessary.
A run interval represents the specific number of repetitions required to trigger and execute the serverless function.
\toolName provides a fine-grained and high-accuracy guaranteed stopping criterion to determine if the performance result set of the current test is sufficiently accurate and reliable to terminate the execution and collection of further repeated runs.
If the performance result set of the current test passes this stopping criterion, it is considered accurate and reliable to represent the actual performance of the serverless function. Otherwise, developers are notified to collect performance results of the serverless function by requesting more repeated runs from another run interval.

Fig.~\ref{fig:overview} gives the workflow of \toolName. 
First, developers input a set of end-to-end response latencies generated by the serverless function to be evaluated. These performance results are denoted as $S_{cur}$ (\textit{Step 1}).
Second, \toolName calculates the required values (e.g., the non-parametric confidence interval for the median~\cite{laaber2019software}) and then performs the accuracy check (\textit{Steps 2 and 3}). Note that the non-parametric confidence interval allows us to estimate percentile performance without relying on strong assumptions about the data distribution. We also provide three specific methods for calculating these non-parametric confidence intervals in Section~\ref{sec:scopeimplementation}.
To ensure the reliability of the performance result acquired with our approach, \toolName also conducts the same performance accuracy analysis for the performance result set obtained from the previous run intervals, denoted as $S_{pre}$. $S_{pre}$ is obtained by removing performance data produced by the last run interval from the end of the $S_{cur}$ (\textit{Step 1*}).
\toolName then calculates the required values for $S_{pre}$ and conducts the accuracy check (\textit{Step 2*} and \textit{Step 3*}). 
Next, \toolName performs the consistency check of the stopping criterion to check whether the returned results of \textit{Step 3} and \textit{Step 3*} are both true (\textit{Step 4}), i.e.,  satisfying the accuracy check. 
If yes, \toolName can give information that the performance result set of the current test $S_{cur}$ is available to represent the actual serverless function performance, and no further repeated runs are needed. 
Otherwise, \toolName waits for developers to collect and provide new end-to-end response latencies when triggering the serverless function repeatedly (i.e., for the number of repetitions specified by the run interval). These new performance data are denoted as $data_{new}$ and added to $S_{cur}$, resulting in an updated set of performance results for evaluation.
\toolName conducts a new round of performance testing using the same processing steps.

\subsection{Stopping criterion of \toolName}\label{stoppingcriterion}


Our stopping criterion includes the accuracy check and the consistency check. The accuracy check is designed to ensure the high accuracy of the obtained performance result, while the consistency check aims to alleviate the possible influence of result fluctuations and ensure result stability. 



For the accuracy check, we assume that a specific performance can be accurate if there exists a desired non-parametric confidence interval for that performance. In our work, the presence of such a desired confidence interval for a given performance is denoted as $dci$.
Specifically, \toolName checks if the confidence intervals for specified metrics (at the given confidence level $cl\%$), calculated from the performance result set of the current test ($S_{cur}$), lie within the corresponding $r\%$ error margins of the observed true value for metrics. If it is yes, there exists the desired confidence intervals for specified metrics.
For example, \toolName determines whether the calculated \textit{95\% confidence interval for the 50\textit{th} percentile} lies within a 1\% error margin of the observed true 50\textit{th} percentile performance. In other words, \toolName determines whether $dci_{50th}$ exists.
Adopting such an accuracy check aims to find the desired and accurate non-parametric confidence interval (CI), where the obtained empirical value of a specific metric (e.g., 50\textit{th} percentile) differs from its observed true performance by no more than the $r\%$ error at a given confidence level. 
To improve the overall accuracy of the performance result acquired with \toolName, \toolName extends the same performance accuracy check to most performance of $S_{cur}$. Most performance of a performance distribution generally includes performance results from the 25\textit{th} percentile to the 75\textit{th} percentile~\cite{maricq2018taming, zhao2021understanding-34, uta2020big}. Moreover, the 50\textit{th} percentile performance represents the midpoint of the performance distribution and plays an important role in performance analysis~\cite{maricq2018taming, uta2020big}. 
Thus, our accuracy check transforms into whether the CIs for the 25\textit{th}, 50\textit{th}, and 75\textit{th} percentiles, calculated from $S_{cur}$, fall within the $r\%$ error margin of their respective observed true percentile performance, referred to \textit{Step 3} in Fig.~\ref{fig:overview}. In other words, \toolName checks whether $dci_{25th}$, $dci_{50th}$, and $dci_{75th}$ exist.
If it is satisfied, $S_{cur}$ is considered accurate. Thus, the accuracy check is formulated as (\ref{eq:accuracycheck}).

\begin{align}
    \text{AccuracyCheck} = 
\begin{cases} 
\text{True} & \text{if } (dci_{25th} = \text{True}) \land (dci_{50th} = \text{True}) \land (dci_{75th} = \text{True}) \\
\text{False} & \text{otherwise}
\end{cases}
    \label{eq:accuracycheck}
\end{align}

Our approach primarily relies on CI since it can achieve a robust analysis in the face of random performance fluctuations~\cite{uta2020big}. Moreover, CI has been established as a valuable metric in performance engineering~\cite{maricq2018taming, uta2020big, laaber2019software}. Consequently, we are based on CI to design and customize our accuracy check for performance testing of serverless computing in our work.

For the consistency check, \toolName checks whether both $S_{cur}$ and the performance result set obtained from the previous run intervals ($S_{pre}$) are accurate.
This is because potential performance fluctuations could make $S_{cur}$ get a temporary result of meeting the accuracy check. However, performance data might be unstable and thus yield unreliable testing results.
Thus, we take the accuracy of $S_{pre}$ into account. 
Our consistency check is to ensure the accuracy of both $S_{cur}$ and $S_{pre}$, formulated as (\ref{eq:consistencycheck}).

\begin{align}
    \text{ConsistencyCheck} = 
\begin{cases} 
\text{True} & \text{if } (AccuracyCheck_{S_{cur}} = \text{True}) \land (AccuracyCheck_{S_{pre}} = \text{True})  \\
\text{False} & \text{otherwise}
\end{cases}
    \label{eq:consistencycheck}
\end{align}

Therefore, we deem $S_{cur}$ to be stable when $S_{pre}$ obtained from previous run intervals also exhibits most of its performance to be accurate. 
That is to say, for both $S_{cur}$ and $S_{pre}$, the calculated CIs for the 25\textit{th}, 50\textit{th}, and 75\textit{th} percentiles all lie within the $r\%$ error margin of the observed true 25\textit{th} percentile performance, 50\textit{th} percentile performance, and 75\textit{th} percentile performance, respectively. This process refers to \textit{Step 4} in Fig.~\ref{fig:overview}.



\subsection{Implementations of \toolName}\label{sec:scopeimplementation}

\toolName relies on the accuracy check, which compares the range of the calculated non-parametric CI with the $r\%$ error margin.
Particularly, we consider non-parametric methods to calculate the required CIs in our approach design. 
This is because the performance of most serverless functions follows a non-normal distribution in both cold and warm starts~\cite{wen2023revisiting}.
Non-parametric methods are suitable for the performance analysis of serverless functions. 
Moreover, they work for the performance data with a normal distribution~\cite{maricq2018taming}. 
Note that in non-parametric analysis, the probability distribution of performance data is unknown, and fewer assumptions are required. Therefore, the metrics related to mean and standard deviation are rarely used because they are not robust when dealing with non-normal distribution~\cite{laaber2019software}.
In our approach, we choose percentile performance metrics because they do not depend on specific distribution assumptions. Calculating CIs for percentile performance also follows a non-parametric approach because it is a way to perform statistical analysis without assuming that the data follow any particular distribution.
In Fig.~\ref{fig:overview}, we support three mainstream non-parametric calculations of CIs for percentile performance (\textit{Step 2} and \textit{Step 2*}). Thus, we obtain the following three implementations of \toolName to provide the user with flexibility.


$\bullet$ \textbf{\textit{\toolName 1}}: We use the general method~\cite{le2010performance, uta2020big, maricq2018taming} to calculate CIs for the percentile. The method involves sorting performance results in ascending order and then calculating two index values based on the data size, the given percentile, and the desired confidence level. Values at these two locations are the lower and upper bounds of the CI for the specified percentile and confidence level. 





$\bullet$ \textbf{\textit{\toolName 2}}: We use the basic bootstrapping method~\cite{basicBT} based on resampling technique to calculate CIs. In this method, a set of $n$ performance data is randomly sampled with replacement to construct a new set. This selection process is repeated $n$ times to form one resample. The resampling process is then repeated $c$ times (at least 1,000 times~\cite{he2021performance}) to generate $c$ resample sets. For each resample, the performance at a specific percentile is calculated, and the resulting values are sorted. Finally, the lower and upper bounds of the CI are determined based on the sorted $c$ values and a given confidence level $cl\%$.





$\bullet$ \textbf{\textit{\toolName 3}}: 
We use the block bootstrapping method~\cite{blockBT} to calculate CIs. Unlike the basic bootstrapping method, in the block bootstrapping method, the data selection of a round resample becomes the selection and combination of the block data with continuous performance results. We apply the automated selection of block size used in this work~\cite{he2021performance} to this method. Conducting $c$ times of resamples will obtain the total of $c$ values about the percentile performance. These values are still sorted, and the lower and upper bounds of the CI for this percentile are generated with a given confidence level $cl\%$.







\subsection{An illustrating example of applying \toolName}

\toolName ensures scalability without requiring additional manual efforts or modifications to the existing serverless platforms. It can be seamlessly integrated as an external service or auxiliary analysis tool to assess performance results. Serverless functions, including complex tasks, are triggered and executed on original serverless platforms. Developers who employ \toolName capture and collect performance data from the serverless platform and input it into \toolName. Then, \toolName automatically analyzes performance data and provides insights into the need for additional repetition runs. This flexibility makes \toolName highly scalable for various serverless functions.

To illustrate how \toolName is used, we provide a real-world example of assessing the performance of a serverless function. The serverless function is from the dataset described in Section~\ref{sec:dataset} of the experimental evaluation. We use \textit{Func43} from the work~\cite{mahgoub2022orion-36} executed on AWS Lambda.
First, the developer gives a set of performance results of a serverless function executed on its serverless platform (\textit{Step 1}), e.g., 180 performance data points regarding the function \textit{Func43}. 
Then, \toolName calculates the corresponding CIs for the 25\textit{th}, 50\textit{th}, and 75\textit{th} percentiles (\textit{Step 2}). 
\textit{Step 3} checks whether these CIs fall within the $r\%$ (e.g., 1\%) error margin of respective observed true percentile performance, i.e., actual 25\textit{th}, 50\textit{th}, and 75\textit{th} percentile performance in performance distribution with 180 data points.
The result shows that these CIs satisfy the accuracy check.
At the same time, \toolName uses a set of performance results from \textit{Step 1} to get the performance result set from previous run intervals, as outlined in \textit{Step 1*}. If the number of repetitions of the run interval is set to 5, the performance result set from previous run intervals is composed of the first 175 data points, excluding the last 5 data points.
Similarly, \toolName calculates and checks the accuracy of the newly calculated CIs for the 25\textit{th}, 50\textit{th}, and 75\textit{th} percentiles (\textit{Step 2* and Step 3*}). The result shows that these CIs obtained from 175 data points satisfy the accuracy check.
Both \textit{Step 3} and \textit{Step 3*} satisfy the corresponding accuracy checks (\textit{Step 4}), and then the performance result set given in \textit{Step 1} is reported as the final performance of \textit{Func43}. 

\section{Experimental evaluation}\label{sec:evaluation}

\subsection{Research questions}

We explore the following research questions:




\textbf{RQ1}: \textit{How general effective is \toolName for serverless functions compared to the state-of-the-art techniques?} This RQ aims to compare the general effectiveness of \toolName and state-of-the-art performance testing techniques for cloud computing on serverless functions.


\textbf{RQ2}: \textit{How well do \toolName and state-of-the-art techniques apply to serverless functions under varying parameters?} The performance testing techniques aim to design the stopping criterion to determine whether the tested performance result set accurately reflects real performance and whether additional repeated runs from another run interval are necessary. The number of repetitions within the run interval represents the amount of new performance data added in each round of testing. 
Based on this, we investigate the effectiveness of \toolName and state-of-the-art techniques under different constraints of the stopping criterion and different numbers of repetitions within the run interval.

\textbf{RQ3}: \textit{How flexible and effective is \toolName compared to the strategy of setting a fixed number of repetitions for all evaluated serverless functions?} As commonly used in previous studies, a fixed number of repetitions is applied for all serverless functions. This RQ aims to investigate the flexibility and effectiveness of \toolName compared to the strategy of setting a fixed number of repetitions for all evaluated serverless functions.


\subsection{Baselines}\label{sec:baselines}
To answer RQ1 and RQ2, we select state-of-the-art performance testing techniques for comparison. As serverless computing is a cloud computing paradigm, a straightforward idea is to adopt existing performance testing techniques designed for cloud applications to serverless functions.
Thus, we consider two state-of-the-art performance testing techniques designed for cloud applications: \compareOne~\cite{he2019statistics} and \compareTwo~\cite{he2021performance}. These approaches are non-parametric and have been evaluated to be superior to other techniques for cloud applications~\cite{he2019statistics, he2021performance}, e.g., detecting repetitiveness of performance data and analyzing coefficient of variation of performance data.
The cloud applications that they test are developed based on IaaS~\cite{bhardwaj2010cloud}, a traditional cloud computing pattern that allows developers to lease resources and configure and manage the infrastructure. 
In addition, we also consider another non-parametric method, \compareThree~\cite{maricq2018taming}, for cloud environments. \compareThree estimates the required number of repetitions for an experiment.

The stopping criteria of \compareOne~\cite{he2019statistics} and \compareTwo~\cite{he2021performance} rely on the stability assessment of performance distributions. 
\compareOne and \compareTwo respectively compare the distribution similarity and changes in a performance metric. Particularly, \compareOne needs to set an objective probability $p_0$ (e.g., 90\%) to represent accuracy requirements. \compareTwo needs to specify a maximum allowed percentage error $e_{0}\%$ (e.g., 3\%). 
The stopping criterion of  \compareThree~\cite{maricq2018taming} relies on the correctness assessment of CIs. It uses resampling without replacement and CIs for the median to determine whether the mean CIs fall within a desired error bound $e_{0}\%$ (e.g., 3\%).



\subsection{Dataset}\label{sec:dataset}
To evaluate the effectiveness of performance testing techniques, we use a representative dataset that has recently been made publicly available for serverless function performance analysis~\cite{wen2023revisiting}. This dataset comprises 65 serverless functions that have been meticulously sourced from peer-reviewed papers published in top-tier academic venues spanning the period from 2014 (the year that serverless computing started to be popular~\cite{JonasCoRR2019new, Wen21challenges}) to 2022. The size of our dataset (i.e., 65 serverless functions) is comparable to and even larger than those used in previous studies of serverless computing performance~\cite{WangATC2018, wen2023FaaSlight, wen2021characterizing, yu2020characterizing, wen2021measurement, zhang2019video}.

The 65 serverless functions span a wide range of task types, ensuring a diverse representation of workloads. Specifically, it covers 25 distinct task categories, such as mathematical operations, image processing, face detection, graph network analysis, video processing, and natural language processing. This broad task variety provides a comprehensive view of real-world serverless computing workloads. Furthermore, the dataset covers widely-used benchmarks in the serverless computing research community~\cite{yu2020characterizing, kim2019functionbench, maissen2020faasdom} and the industry~\cite{awsserverlessrepo}, e.g., \textit{ServerlessBench}~\cite{ServerlessBench}, \textit{FunctionBench}~\cite{FunctionBench}, and \textit{FaaSDom}~\cite{FaaSDom}. In terms of serverless platforms, the dataset focuses on AWS Lambda and Google Cloud Functions, which are the most prevalent public serverless platforms to study~\cite{serverlesscommunitysurvey1, eismann2021state, hassan2021survey}. For programming languages, the dataset includes functions written in Python and JavaScript, which are the dominant programming languages in serverless computing~\cite{serverlesscommunitysurvey, serverlesscommunitysurvey1, eismann2021state, eskandani2021wonderless}.

\subsection{Experimental setup}\label{sec:experimentalsetup}

\noindent\textbf{Execution configurations of serverless functions.} 
We execute 65 serverless functions using the original function configurations and serverless platforms specified in the work~\cite{wen2023revisiting}. 
If a configuration is not provided, we use the default configuration of the platform.
At the time of our study, AWS Lambda uses a default memory size of 128 MB~\cite{AWSmemory} and a timeout of 3 seconds~\cite{AWStimeout}, while Google Cloud Functions adopts a default memory size of 256 MB~\cite{Googlememory} and a timeout of 60 seconds~\cite{Goolgetimeout}.
If the configured memory size or timeout time is insufficient to support executions, we increase the value of these parameters and test the function again to ensure successful execution.

We repeatedly invoke the serverless function to produce a series of performance data, which are input to performance testing techniques to check when the repeated runs can be stopped. If the stopping criterion is not satisfied, we start the next run interval to generate more performance data. Since our goal is to evaluate the effectiveness of performance testing techniques in the context of serverless computing, by default, we set the number of repetitions of the run interval to five repetitions, an established number of repetitions commonly used in the experimental setting of serverless computing papers~\cite{mvondo2021ofc, ao2022faasnap, jangda2019formal-151}.
This small number allows us to obtain a fine-grained stop location and reduce the unnecessary overhead of running the function.
In RQ2, we investigate the effect of the size of this number.
We evaluate the effectiveness of performance testing techniques using cold-start and warm-start performance of the serverless functions. 
The cold-start performance is obtained by invoking the function after the resources generated by previous invocations have been released. 
The warm-start performance is obtained by invoking the function before releasing the resources from previous invocations. 
We use a half-hour invocation frequency for cold-start performance and a five-second frequency for warm-start performance, both after the previous invocation, as they can ensure the serverless function experiences cold and warm starts.
For performance testing techniques, we use the same performance data of the functions to fairly compare their effectiveness.

\noindent\textbf{Parameter configurations.} 
In experiment evaluation, we will explore the supported three versions of \toolName (i.e., \textit{\toolName 1}, \textit{\toolName 2}, and \textit{\toolName 3}) to demonstrate the flexibility and applicability of our approach across different non-parametric confidence interval calculations for percentile performance. For the comparison of evaluation results, we use the version of \toolName (e.g., \textit{\toolName 1}) that is the most effective.

To answer RQ1, the confidence level $cl\%$ and the error margin $r\%$ in \toolName use by default the values that have been widely adopted by previous work on performance analysis~\cite{maricq2018taming, uta2020big}.
$cl\%$ and $r\%$ are set to 95\% and 1\%, respectively.
The resample times in \textit{\toolName 2} and \textit{\toolName 3} are by default set to 1,000, consistent with previous work~\cite{he2021performance}.
For \compareOne and \compareTwo, we use the default settings provided in their open-sourced code. 
For \compareOne, we use the default objective probability ($p_{0}$) of 90\%, i.e., expecting the accuracy of testing results to be at least 90\%. For \compareTwo, we use the maximum allowed percentage error ($e_{0}\%$) of 3\% for the median performance and the confidence level of 95\%.
For \compareThree, we use the same error $e_{0}\%$ of 3\% for the CIs for the median performance and the confidence level of 95\%, as in \compareTwo.

To answer RQ2, we adjust key parameters. (1) First, we adjust $r\%$ of \toolName to 5\%, 4\%, 3\%, 2\%, and 1\% to investigate the characteristics of the stopping criterion. We also adjust the key parameters for the stopping criterion of the state-of-the-art techniques: $p_{0}$ of \compareOne and $e_{0}\%$ of \compareTwo and \compareThree. 
We set $p_{0}$ to values of 90\%, 92\%, 94\%, 96\%, and 98\%, where 92\%, 94\%, 96\%, and 98\% are stricter constraints that have not been used in previous evaluations of \compareOne~\cite{he2019statistics}. 
We set $e_{0}\%$ to values of 5\%, 4\%, 3\%, 2\%, and 1\%, where 2\% and 1\% are also stricter constraints not used in previous evaluations of \compareTwo~\cite{he2021performance, maricq2018taming}.
(2) Second, we adjust the number of repetitions within the run interval to 3, 4, 5, 10, and 20, which are commonly used in serverless computing papers~\cite{perron2020starling-26, fuerst2022locality-32, mvondo2021ofc, wang2019replayable, AkkusATC18-131, ao2022faasnap, jangda2019formal-151} for the total experimental repetitions. Serverless computing scenario generally uses small repetitions. The possible reason is that serverless functions can be triggered at any time, which makes it convenient to obtain any number of performance results. Thus, in the serverless computing scenario, developers tend to use a small run interval to invoke serverless functions on demand at any time without needing to lease resources in advance.
Taking these values as the number of repetitions of one run interval, we could investigate if adding a cycle of performance data affects testing results and if the repetitions specified in previous work are sufficient to obtain accurate serverless function performance.
We use default configurations for other parameters of \toolName, \compareOne, \compareTwo, and \compareThree.

To answer RQ3, we evaluate the strategy of indiscriminately setting a fixed number of repetitions for all tested serverless functions, in the same way used in prior serverless computing studies. This strategy is different from our compared baselines. We evaluate different values from small to large, including 20~\cite{ustiugov2021benchmarking-161}, 50~\cite{OakesATC18-140}, 100~\cite{kotni2021faastlane-130}, 300~\cite{mahgoub2022orion-36}, and 500~\cite{datta2022alastor}. They had been used in serverless computing papers.

\noindent\textbf{Evaluation strategy and metrics.} 
We use the performance data of the serverless function being repeatedly executed 1,000 times as the ground truth performance for identifying the effectiveness of testing results. To establish the ground truth, we require a relatively large number of performance tests that can capture all potential impacts of the serverless platform. We observe that 1,000 repetitions are the largest number found in existing serverless computing literature~\cite{WangATC2018, li2022faasflow-74, wen2023revisiting}. To confirm whether 1,000 repetitions are sufficient, we execute each serverless function for an additional 500 runs, and the results, discussed in Section~\ref{sec:groundtruth}, show consistent effectiveness compared to 1,500 repetitions. Therefore, 1,000 executions provide a trustworthy ground truth.
We apply performance testing techniques to determine the termination of the repeated runs for the serverless function, i.e., the stop location. The performance data tested in the stop location is deemed to be the accurate performance distribution of this function acquired with performance testing techniques. 
We compare this distribution with the performance distribution of the ground truth of this function using the following evaluation metrics.

$\bullet$ \textit{Accuracy:} He et al.~\cite{he2019statistics} define the accuracy of performance testing results as the similarity between the performance distribution acquired with performance testing techniques and the corresponding ground truth distribution.
The value of the similarity metric ranges from 0 to 100\%, where 100\% indicates the same distribution. 
We calculate the mean accuracy of the testing results obtained for all tested functions.

$\bullet$ \textit{Reliability:} Previous work~\cite{uta2020big} defines the reliability of the obtained performance result as whether its specific percentile performance is accurate. 
When a percentile performance obtained from the performance distribution acquired with performance testing techniques falls within the 95\% confidence interval for this percentile obtained from the corresponding ground truth distribution, it indicates a 95\% probability that this percentile performance is reliable~\cite{uta2020big}. 
It also indicates that \textit{the performance testing techniques enable the tested serverless function to get this reliable percentile performance.}
This previous work~\cite{uta2020big} has focused on the performance at the 50\textit{th} and 90\textit{th} percentiles to assess the result reliability. 
To obtain more comprehensive results, we investigate the reliability of testing results from different performance perspectives by extending the percentiles to include the 25\textit{th}, 50\textit{th}, 75\textit{th}, and 90\textit{th} percentiles. 
We respectively calculate the percentage of the serverless functions with reliable performance at different percentiles.


\noindent\textbf{Experimental environment.}
We implemented and ran \toolName, \compareOne, \compareTwo, \compareThree and the invocation scripts for serverless functions on an Ubuntu 18.04.4 LTS server with an Intel Xeon (R) 4-core processor and 24GiB of memory. 

\section{Results}

\subsection{RQ1: General Effectiveness of \toolName}

\begin{table}[t]
\footnotesize
    \caption{RQ1: The results of performance testing for 65 serverless functions using \toolName, \compareOne, \compareTwo, and \compareThree.}
    \label{tab:rq1}
    \begin{tabular}{l|p{2cm}|p{1.8cm}|p{1.8cm}|p{1.8cm}|p{1.8cm}}
        \hline
         \begin{tabular}[c]{@{}l@{}} \end{tabular}  &  \tabincell{c}{\textbf{Mean accuracy}} & \multicolumn{4}{c} { \tabincell{c}{\textbf{\textit{\#Functions}: reliability at 25\textit{th}/50\textit{th}/75\textit{th}/90\textit{th} percentile}}} 
         \\ \hline
         
        \tabincell{c}{ \textit{\toolName 1}  -  cold start} & 97.39\%	& 89.23\%	& 93.85\%	& 90.77\% &	92.31\% \\

         \hline 
        \tabincell{c}{ \textit{\toolName 1} - warm start} & 97.12\% &	86.15\%	& 92.31\% &	93.85\% &	89.23\%  \\ \hline 
        \textbf{\textit{\toolName 1} (Mean)} & \textbf{97.25\%} &	\textbf{87.69\%}	& \textbf{93.08\%} &	\textbf{92.31\%}	& \textbf{90.77\%} \\ 
        \hline
        \hline 
        \tabincell{c}{ \textit{\toolName 2} - cold} & 95.94\% &	84.62\% &	86.15\% &	73.85\% &	73.85\%   \\ \hline 
        \tabincell{c}{ \textit{\toolName 2} - warm} &  94.19\%	& 75.38\% &	80.00\% &	75.38\% &	84.62\% \\ \hline
        \textbf{\toolName 2 (Mean)} & \textbf{95.07\%} &	\textbf{80.00\%} &	\textbf{83.08\%} &	\textbf{74.62\%} &	\textbf{79.23\%} \\ \hline
        \hline 
        \tabincell{c}{ \textit{\toolName 3} - cold start}  & 95.05\% &	84.62\% &	78.46\% &	70.77\% &	67.69\% \\ \hline 
        \tabincell{c}{ \textit{\toolName 3} - warm start}  &  93.22\% &	66.15\% &	78.46\% &	81.54\% &	83.08\%  \\ \hline
        \textbf{\textit{\toolName 3} (Mean)} & \textbf{94.13\%}	& \textbf{75.38\%} &	\textbf{78.46\%} &	\textbf{76.15\%} &	\textbf{75.38\%} \\ \hline
        \hline
        \compareOne \ - cold & 61.12\% & 12.31\% &	10.77\% &	18.46\% &	29.23\%  \\ \hline 
        \compareOne \ - warm & 43.45\%	& 16.92\% &	20.00\% &	18.46\% &	12.31\%  \\ \hline 
        \textbf{\compareOne (Mean)} & \textbf{52.28\%}	& \textbf{14.62\%} &	\textbf{15.38\%} &	\textbf{18.46\%} &	\textbf{20.77\%} \\ \hline
        \hline
        \compareTwo \ - cold start  & 69.38\%	& 26.15\% &	21.54\% &	20.00\% &	27.69\% \\ \hline 
        \compareTwo \ - warm start & 57.47\% &	24.62\% &	23.08\% &	23.08\% &	21.54\%  \\ \hline 
        \textbf{\compareTwo (Mean)} & \textbf{63.42\%} &	\textbf{25.38\%} &	\textbf{22.31\%} &	\textbf{21.54\%} &	\textbf{24.62\%}	\\ \hline
        \hline
        \compareThree \ - cold start  & 68.93\%	& 24.62\% &	21.54\% &	23.08\% &	16.92\%\\ 
        \hline 
        \compareThree \ - warm start & 51.91\% &	13.85\%	 & 16.92\%	& 29.23\% &	21.54\% \\ \hline 
        \textbf{\compareThree (Mean)} & \textbf{60.42\%} &	\textbf{19.23\%} &	\textbf{19.23\%} &	\textbf{26.15\%} &	\textbf{19.23\%}	\\ \hline
    \end{tabular}
\end{table}

This section explores the general effectiveness of \toolName compared to \compareOne, \compareTwo, and \compareThree when testing serverless function performance. Results show that \toolName is highly effective. Table~\ref{tab:rq1} shows the results of three variants of \toolName and state-of-the-art techniques in cold-start and warm-start performance testing for serverless functions. On average, the mean accuracy obtained by \textit{\toolName 1}, \textit{\toolName 2}, and \textit{\toolName 3} is 97.25\%, 95.07\%, and 94.13\%, respectively. \compareOne, \compareTwo, and \compareThree provide testing results with 52.28\%, 63.42\%, and 60.42\% mean accuracy, respectively. Compared to \compareOne, \compareTwo, and \compareThree, \toolName can improve the mean accuracy by 44.97, 33.83, and 36.83 percentage points, respectively. Moreover, \toolName outperforms \compareOne, \compareTwo, and \compareThree on all serverless functions that we consider, thus indicating the feasibility and effectiveness of \toolName.



For the reliability, \textit{\toolName 1} provides the reliable 25\textit{th}, 50\textit{th}, 75\textit{th}, and 90\textit{th} percentile performance for 87.69\%, 93.08\%, 92.31\%, and 90.77\% of the serverless functions, respectively. 
\textit{\toolName 2} provides the reliable 25\textit{th}, 50\textit{th}, 75\textit{th}, and 90\textit{th} percentile performance for 80.00\%, 83.08\%, 74.62\%, and 79.23\% of the functions, respectively. 
\textit{\toolName 3} provides similar reliability to \textit{\toolName 2}.
For \compareOne, on average, it enables 14.62\%, 15.38\%, 18.46\%, and 20.77\% of the functions to get the reliable 25\textit{th}, 50\textit{th}, 75\textit{th}, and 90\textit{th} percentile performance, respectively. 
Compared to \compareOne, \toolName provides the reliable 25\textit{th}, 50\textit{th}, 75\textit{th}, and 90\textit{th} percentile performance for an additional 87.69\% - 14.62\% = 73.07\%, 93.08\% - 15.38\% = 77.70\%, 92.31\% - 18.46\% = 73.85\%, and 90.77\% - 20.77\% = 70.00\% of the functions, respectively.
For \compareTwo, it provides the reliable 25\textit{th}, 50\textit{th}, 75\textit{th}, and 90\textit{th} percentile performance for 25.38\%, 22.31\%, 21.54\%, and 24.62\% of the functions, respectively.
Compared to \compareTwo, \toolName provides the reliable 25\textit{th}, 50\textit{th}, 75\textit{th}, and 90\textit{th} percentile performance for an additional 87.69\% - 25.38\% = 62.31\%, 93.08\% - 22.31\% = 70.77\%, 92.31\% - 21.54\% = 70.77\%, and 90.77\% - 24.62\% = 66.15\% of the functions, respectively.
For \compareThree, it provides the reliable 25\textit{th}, 50\textit{th}, 75\textit{th}, and 90\textit{th} percentile performance for 19.23\%, 19.23\%, 26.15\%, and 19.23\% of the functions, respectively.
Compared to \compareThree, \toolName provides the reliable 25\textit{th}, 50\textit{th}, 75\textit{th}, and 90\textit{th} percentile performance for an additional 87.69\% - 19.23\% = 68.46\%, 93.08\% - 19.23\% = 73.85\%, 92.31\% - 26.15\% = 66.16\%, and 90.77\% - 19.23\% = 71.54\% of the functions, respectively.
These results show that most of the testing results produced with \toolName are accurate and reliable.




For three implementations of \toolName, we summarize the following points. 
(1) \textit{\toolName 1} outperforms the other two, indicating that the CI calculation method of \textit{\toolName 1} has high accuracy on performance testing for serverless functions. 
This could be because the other two implementations use bootstrapping methods, which adopt the resampling strategy. The resampling process may break the original data distribution and introduce potential noises. In contrast, \textit{\toolName 1} is based on the original performance data to calculate CI. 
(2) \textit{\toolName 2} and \textit{\toolName 3} show comparable effectiveness. This indicates that our approach design is insensitive to internal data dependency and has a similar effectiveness for the same type of CI calculation method.
(3) All implementations show comparable effectiveness in both cold and warm starts, indicating the stability of \toolName in performance testing for serverless functions.

\underline{\textit{Result discussion.}} 
We observe that state-of-the-art techniques (\compareOne, \compareTwo, and \compareThree) perform poorly in evaluating serverless functions. This ineffectiveness stems from fundamental differences in their approach: these techniques rely on stability or correctness assessments for specific performance. However, serverless functions, characterized by short durations and small run intervals, make it difficult for these methods to detect significant stability changes or undesired confidence intervals. The minor fluctuations exhibited in performance distributions cause these methods to reach their stopping criterion too early, undermining their ability to accurately assess the need for additional repetitions. This limitation highlights the need for a novel performance testing approach tailored to serverless functions that allows for finer-grained analysis.

However, the assessment strategy of \compareOne, \compareTwo, and \compareThree may be effective for traditional cloud applications or environments~\cite{he2019statistics, he2021performance, maricq2018taming}. This may be because, traditional cloud applications or task executions have long-lived and minute-level duration, and previous work in cloud computing~\cite{bhardwaj2010cloud, he2019statistics, he2021performance} adopted a period of time of runs to constantly invoke them for execution. This time is often several weeks or days, thus yielding a large number of performance results. Therefore, in state-of-the-art techniques, these characteristics of traditional cloud applications or environments lead to significant changes in stability or undesired confidence intervals on performance distributions, making them effective.

Serverless functions have distinctive performance features from traditional cloud applications or environments. Moreover, state-of-the-art techniques are not effective when applied to serverless function performance. In this paper, we present \toolName, which introduces a novel stopping criterion, incorporating accuracy and consistency checks. These checks enable fine-grained analysis and high accuracy guarantees for performance distributions with these characteristics of serverless functions.
This makes \toolName more effective in the performance testing of serverless functions.




\finding{
\toolName provides testing results with 97.25\% accuracy, 44.97, 33.83, and 36.83 percentage points higher than \compareOne, \compareTwo, and \compareThree, respectively.
Moreover, \toolName outperforms all compared baselines on all serverless functions that we consider, thus indicating its feasibility and effectiveness.
}



\subsection{RQ2: Effectiveness under varying parameters}

\begin{figure}[t]
	\centering
 \includegraphics[width=0.5\textwidth]{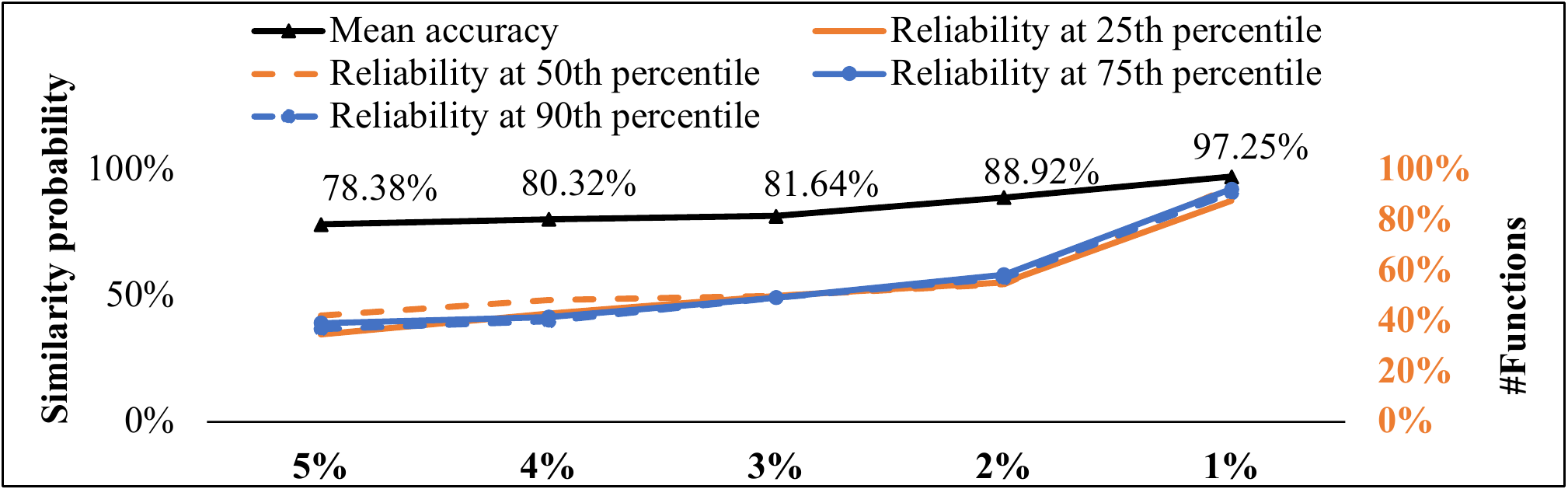}
    \caption{RQ2: Changes in metric values under different constraints $r\%$ for \textbf{\textit{\toolName 1}} (mean results in cold and warm starts for tested functions).}
    \label{fig:SCOPE1_demands}
\end{figure}

\begin{figure}[t]
	\centering
 \includegraphics[width=0.5\textwidth]{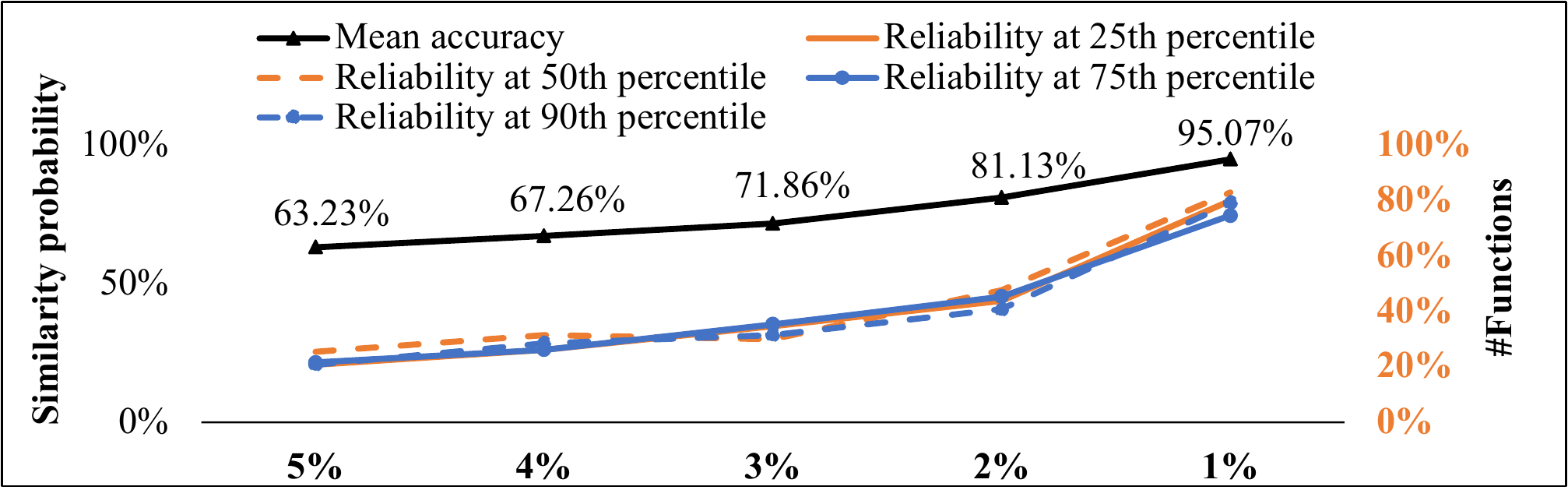}
    \caption{RQ2: Changes in metric values under different constraints $r\%$ for \textbf{\textit{\toolName 2}} (mean results in cold and warm starts for tested functions).}
    \label{fig:SCOPE2_demands}
\end{figure}

\begin{figure}[t]
	\centering
 \includegraphics[width=0.5\textwidth]{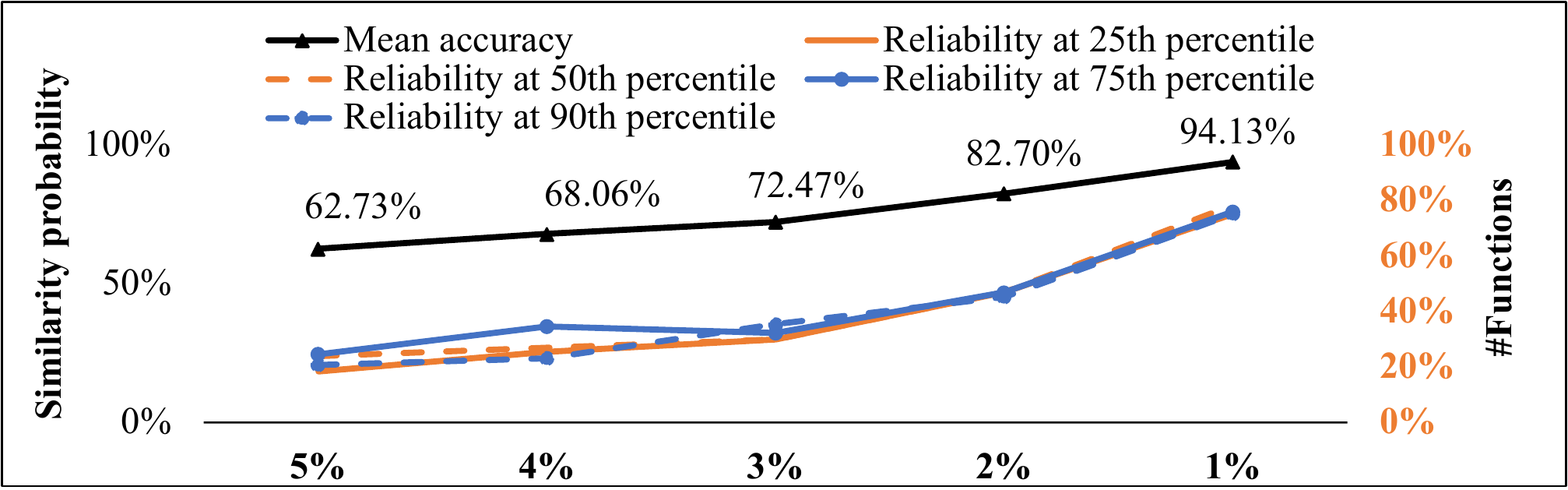}
    \caption{RQ2: Changes in metric values under different constraints $r\%$ for \textbf{\textit{\toolName 3}} (mean results in cold and warm starts for tested functions).}
    \label{fig:SCOPE3_demands}
\end{figure}

\begin{figure}[t]
	\centering
    \includegraphics[width=0.5\textwidth]{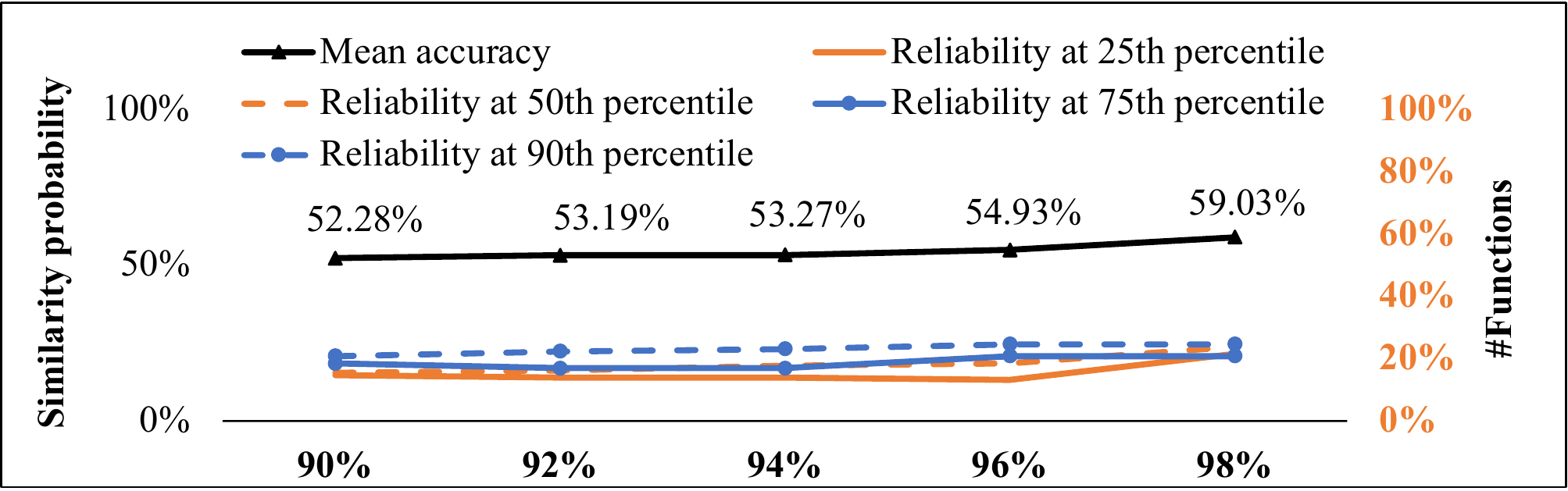}
    \caption{RQ2: Changes in metric values under different $p_{0}$ for \textbf{\compareOne}.}
    \label{fig:FSE_demands}
\end{figure}

\begin{figure}[t]
	\centering
 \includegraphics[width=0.5\textwidth]{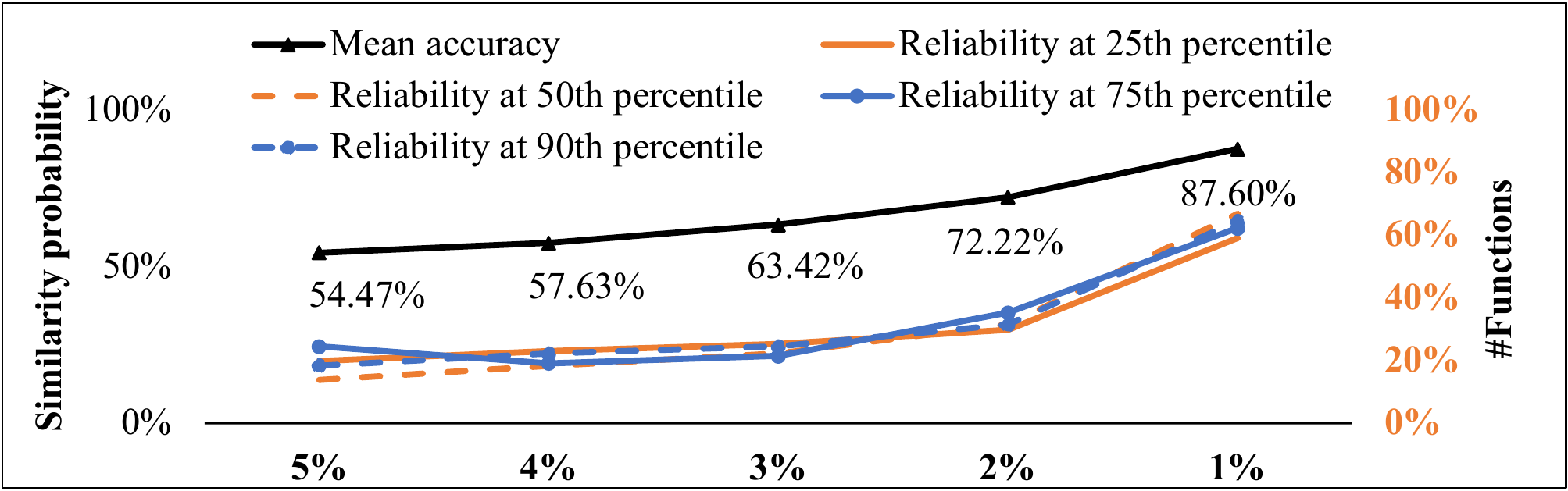}
    \caption{RQ2: Changes in metric values under different $e_{0}\%$ for \textbf{\compareTwo}.}
    \label{fig:ASE_demands}
\end{figure}

\begin{figure}[t]
	\centering
 \includegraphics[width=0.5\textwidth]{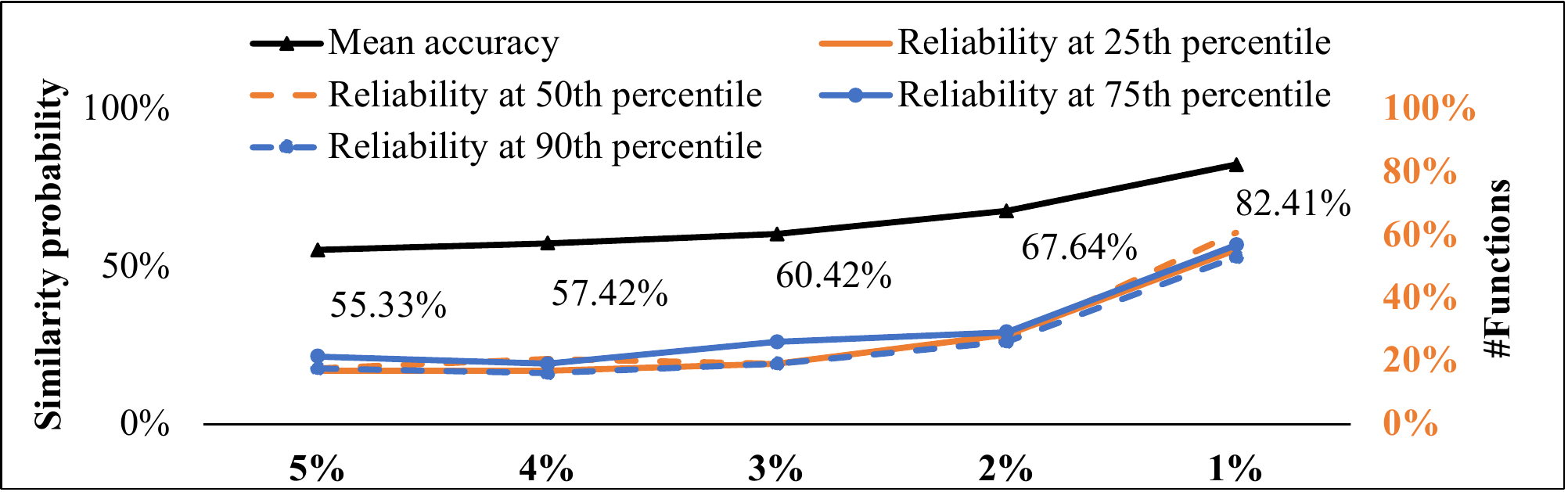}
    \caption{RQ2: Changes in metric values under different $e_{0}\%$ for \textbf{\compareThree}.}
    \label{fig:CONFIRM_demands}
\end{figure}

In this section, we investigate the effectiveness of \toolName and state-of-the-art techniques under varying parameters, including different constraints of the stopping criterion and different numbers of repetitions within the run interval.

We compare the effectiveness of \toolName and state-of-the-art techniques under different constraints of the stopping criterion. 
For \toolName, evaluation metric values have improvements as $r\%$ is limited from 5\% to 1\%. 
\compareOne, \compareTwo, and \compareThree have different sensitivities to the constraint of their stopping criteria. While the effectiveness of \compareOne does not significantly improve with increasing constraints, \compareTwo and \compareThree show improvement in accuracy from 54.47\% to 87.60\% and from 55.33\% to 82.41\%, respectively. However, under the strictest constraint of \compareTwo and \compareThree, the obtained accuracy does not exceed 90\%, which can be obtained by \toolName. Overall, stopping criteria of state-of-the-art techniques may be insufficient for serverless computing.

Figs.~\ref{fig:SCOPE1_demands}, ~\ref{fig:SCOPE2_demands}, and  ~\ref{fig:SCOPE3_demands} show mean changes for three implementations of \toolName under cold and warm starts. 
The mean accuracy improves from 78.38\% to 97.25\% for \textit{\toolName 1}, from 63.23\% to 95.07\% for \textit{\toolName 2}, and from 62.73\% to 94.13\% for \textit{\toolName 3}.
For the reliability, with the use of \textit{\toolName 1}, the proportion of serverless functions that get the reliable 25\textit{th}, 50\textit{th}, 75\textit{th}, and 90\textit{th} percentile performance increases from 34.62\% to 87.69\%, from 42.31\% to 93.08\%, from 39.23\% to 92.31\%, and 36.92\% to 90.77\%, respectively.
\textit{\toolName 2} and \textit{\toolName 3} show the same trends for the reliability as \textit{\toolName 1}.
These results show that the effectiveness of \toolName is influenced by the constraints of the designed stopping criterion. Performance testing for serverless functions requires strict error constraints due to the short duration of the test.
Using a strict error constraint enhances accuracy, but also may increase runtime costs. \toolName offers the advantage of achieving accurate performance without the need for excessive repetitions, thereby avoiding unnecessary runtime costs. In real-world scenarios, \toolName provides developers with the flexibility to adjust the error constraint, allowing them to make desired trade-offs in line with specific requirements.

Fig.~\ref{fig:FSE_demands} illustrates the mean changes in metric values obtained by \compareOne when evaluating the cold-start and warm-start performance of the serverless functions under varying $p_{0}$. The mean accuracy changes from 52.28\% to 59.03\%.
Although \compareOne makes more functions to obtain reliable percentile performance, the improvement is insignificant. Even if $p_{0}$ is constrained to 98\%, only 21.54\%, 23.85\%, 20.77\%, and 24.62\% of the serverless functions can get the reliable 25\textit{th}, 50\textit{th}, 75\textit{th}, and 90\textit{th} percentile performance, respectively.
Thus, for \compareOne, we do not observe significant improvement in evaluation metrics as $p_{0}$ increases from 90\% to 98\%.

Fig.~\ref{fig:ASE_demands} shows the mean changes in metric values obtained by \compareTwo under different $e_{0}\%$. 
The mean accuracy improves from 54.47\% to 87.60\% as $e_{0}\%$ becomes stricter, indicating that using a stricter $e_{0}\%$ can improve the effectiveness of \compareTwo. 
However, even if $e_{0}\%$ is limited to 1\%, the obtained mean accuracy (87.60\%) does not exceed 90\%. Moreover, the percentage of serverless functions with reliable performance is low. For example, for the 25\textit{th} percentile performance, when $e_{0}\%$ = 1\%, \compareTwo enables 59.23\% of the functions to get this reliable performance, while when $e_{0}\%$ = 5\%, only 20.00\% of the functions get this reliable performance.

Fig.~\ref{fig:CONFIRM_demands} shows the mean changes in metric values obtained by \compareThree under different $e_{0}\%$. 
The mean accuracy improves from 55.33\% to 82.41\% as $e_{0}\%$ becomes stricter, indicating that using a stricter $e_{0}\%$ can improve the effectiveness of \compareThree. 
However, even if $e_{0}\%$ is limited to 1\%, the obtained mean accuracy (82.41\%) does not exceed 90\%. Moreover, the percentage of serverless functions with reliable performance is low. For example, for the 90\textit{th} percentile performance, when $e_{0}\%$ = 1\%, \compareThree enables 53.08\% of the functions to get this reliable performance, while when $e_{0}\%$ = 5\%, only 17.69\% of the functions get this reliable performance.

We further investigate the effectiveness of \toolName, \compareOne, \compareTwo, and \compareThree under varying numbers of repetitions of the run interval. 
For \toolName, evaluation results do not show significant changes as the number of repetitions of the run interval increases from 3 to 20. 
\toolName can obtain accuracy results that remain between 96.96\% and 97.53\%, showing a small variation of about 0.50\%.
For \compareOne, \compareTwo, and \compareThree, we observe a positive effect of the run interval on their effectiveness. However, the accuracy never reaches 80\%.

\begin{figure}[t]
	\centering
\includegraphics[width=0.5\textwidth]{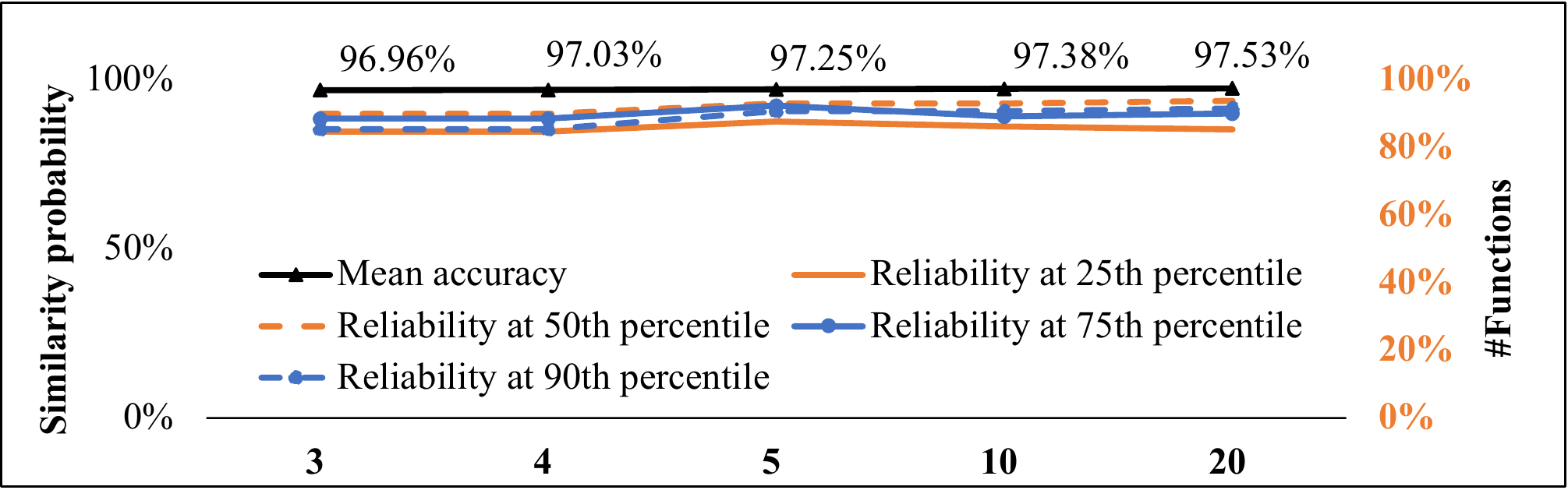}
    \caption{RQ2: Changes in metric values under different numbers of repetitions of the run interval for \textbf{\textit{\toolName 1}} (mean results in cold and warm starts for tested functions).}
    \label{fig:SCOPE1_interval}
\end{figure}
\begin{figure}[t]
	\centering
\includegraphics[width=0.5\textwidth]{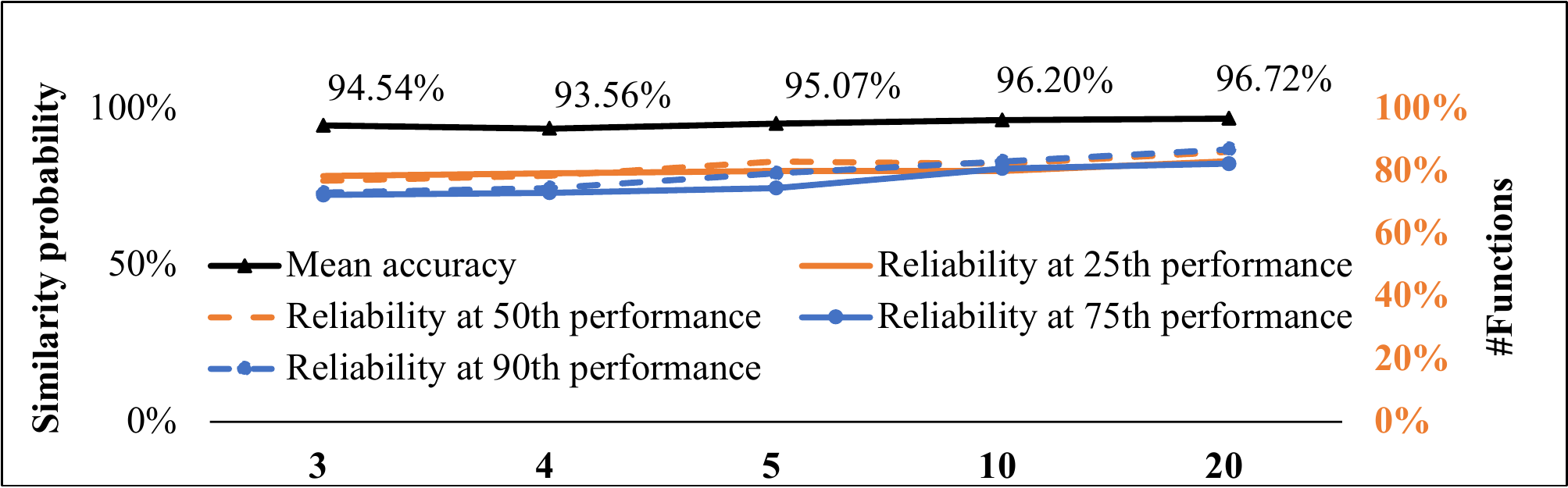}
    \caption{RQ2: Changes in metric values under different numbers of repetitions of the run interval for \textbf{\textit{\toolName 2}} (mean results in cold and warm starts for tested functions).}
    \label{fig:SCOPE2_interval}
\end{figure}
\begin{figure}[t]
	\centering
\includegraphics[width=0.5\textwidth]{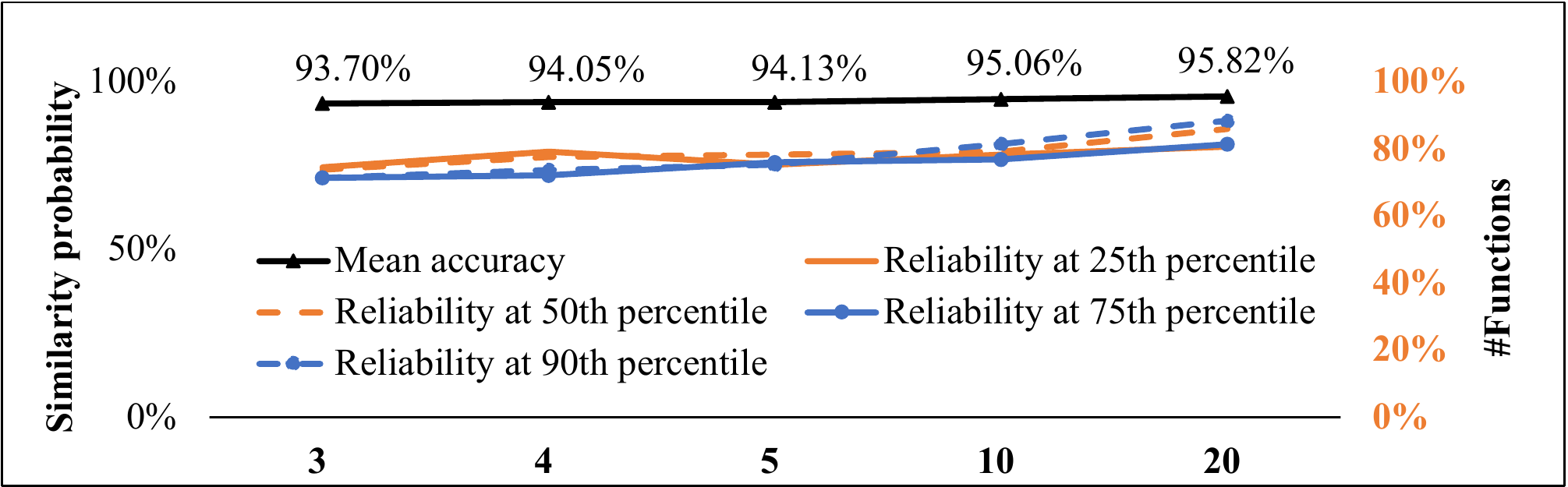}
    \caption{RQ2: Changes in metric values under different numbers of repetitions of the run interval for \textbf{\textit{\toolName 3}} (mean results in cold and warm starts for tested functions).}
    \label{fig:SCOPE3_interval}
\end{figure}

Figs.~\ref{fig:SCOPE1_interval}, ~\ref{fig:SCOPE2_interval}, and ~\ref{fig:SCOPE3_interval} show the mean results obtained by \toolName under cold and warm starts. 
The mean accuracy obtained by \textit{\toolName 1}, \textit{\toolName 2}, and \textit{\toolName 3} ranges from 96.96\% to 97.53\%, from 93.56\% to 96.72\%, and from 93.70\% to 95.82\%, respectively. This indicates a negligible change in accuracy.
The reliability at the percentile performance is also stable. 
As the number of repetitions of the run interval increases, 
\textit{\toolName 1} produces the reliable 25\textit{th}, 50\textit{th}, 75\textit{th}, and 90\textit{th} percentile performance for the functions falling within the following ranges: 84.62\% to 87.69\%, 90.00\% to 93.85\%, 88.46\% to 92.31\%, and 85.38\% to 91.54\%.
\textit{\toolName 2} produces the reliable 25\textit{th}, 50\textit{th}, 75\textit{th}, and 90\textit{th} percentile performance for the functions falling within the following ranges: 78.46\% to 83.08\%, 76.92\% to 86.15\%, 72.31\% to 82.31\%, and 73.08\% to 86.92\%.
The reliability results of \textit{\toolName 3} are highly similar to \textit{\toolName 2}. 
Overall, the effectiveness of \toolName is not affected by the number of repetitions of the run interval. 


\begin{figure}[t]
	\centering
    \includegraphics[width=0.5\textwidth]{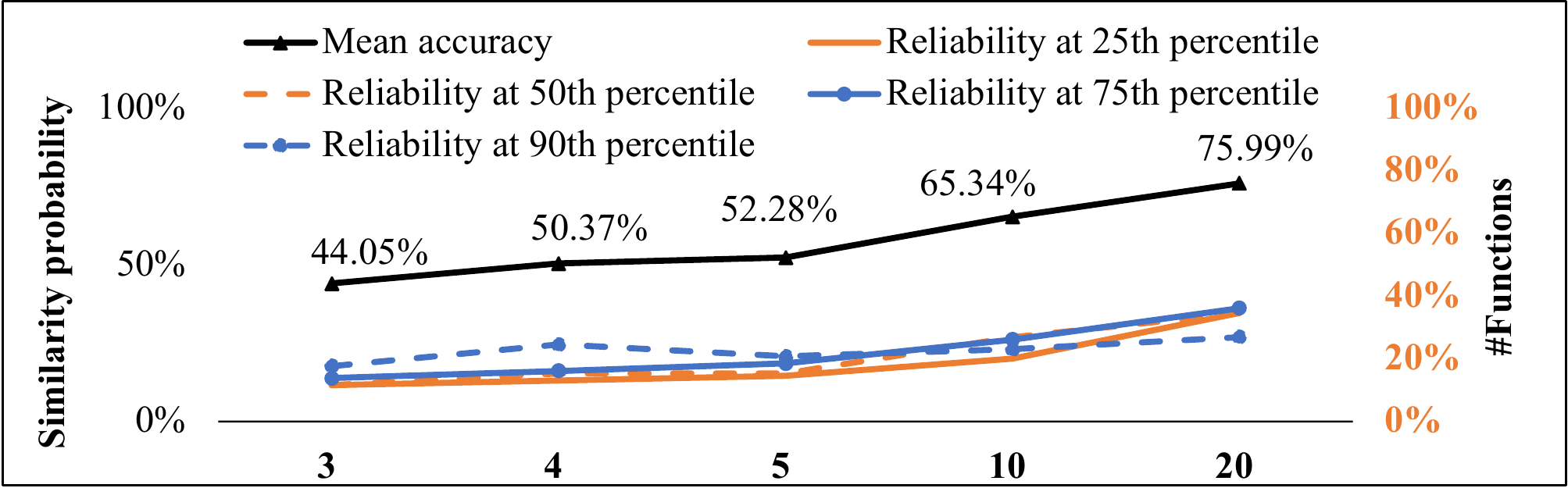}
    \caption{RQ2: Changes in metric values under different numbers of repetitions of the run interval for \textbf{\compareOne}.}
    \label{fig:FSE_interval}
\end{figure}

\begin{figure}[t]
	\centering
    \includegraphics[width=0.5\textwidth]{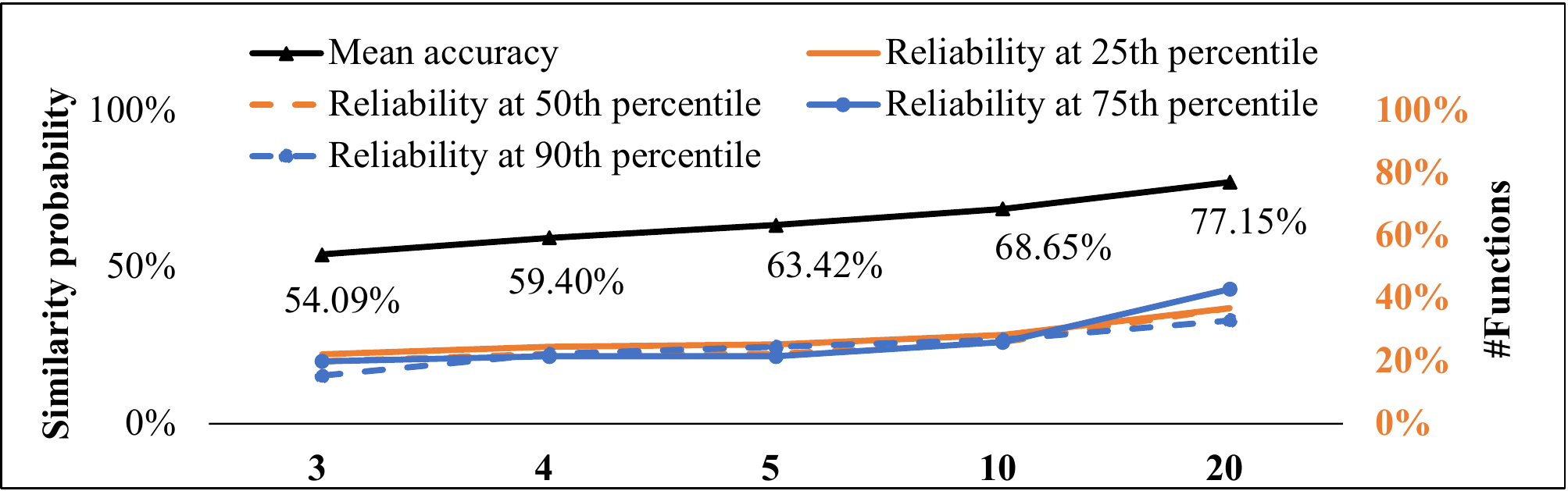}
    \caption{RQ2: Changes in metric values under different numbers of repetitions of the run interval for \textbf{\compareTwo}.}
    \label{fig:ASE_interval}
\end{figure}

\begin{figure}[t]
	\centering
    \includegraphics[width=0.5\textwidth]{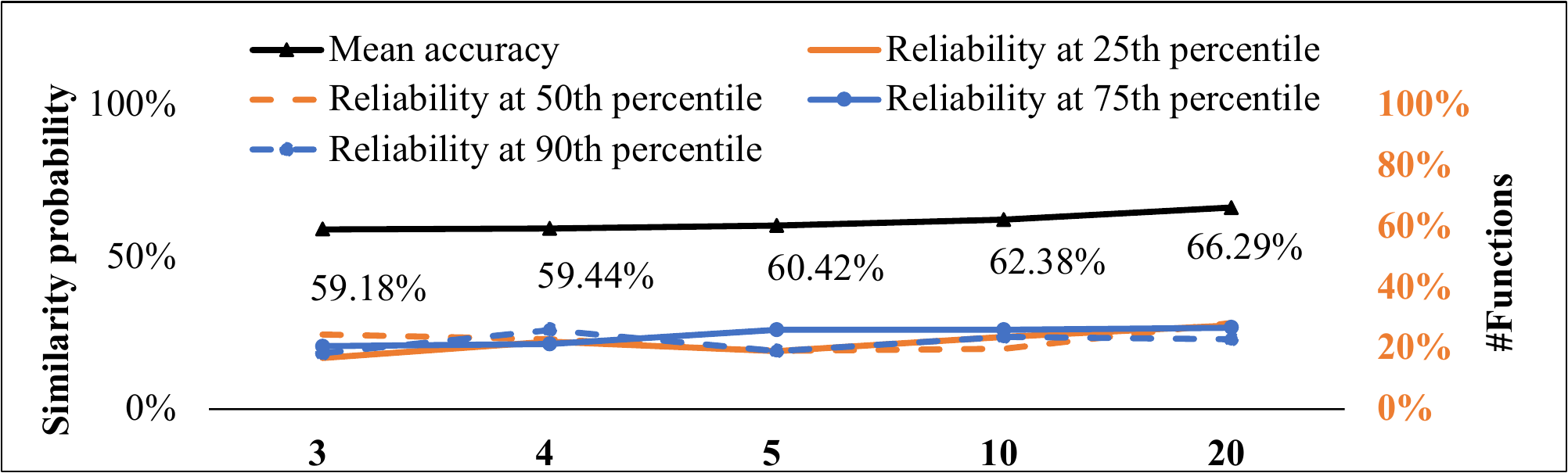}
    \caption{RQ2: Changes in metric values under different numbers of repetitions of the run interval for \textbf{\compareThree}.}
    \label{fig:CONFIRM_interval}
\end{figure}

Fig.~\ref{fig:FSE_interval} shows the mean changes obtained by \compareOne in cold-start and warm-start performance testing. The obtained mean accuracy increases from 44.05\% to 75.99\%, as the number of repetitions of the run interval increases from 3 to 20.
This indicates a positive effect of the run interval on the effectiveness of \compareOne. 
However, while increasing the number of repetitions of the run interval enables more functions to get reliable performance, \compareOne still does not enable over 60\% of the serverless functions to achieve it.

Fig.~\ref{fig:ASE_interval} shows the mean results obtained by \compareTwo in cold-start and warm-start performance testing. As the number of repetitions of the run interval increases from 3 to 20, the obtained mean accuracy increases from 54.09\% to 77.15\%.
This also indicates a positive effect of the run interval on the effectiveness of \compareTwo. 
However, the accuracy of testing results is still no more than 80\%, similar to the results obtained by \compareOne.
In terms of reliability, for example, \compareTwo enables only 36.92\% of the serverless functions to get the reliable 50\textit{th} percentile performance when the number of repetitions of the run interval is set to 20. As a result, 63.08\% of the functions still cannot get this reliable performance.

Fig.~\ref{fig:CONFIRM_interval} shows the mean results obtained by \compareThree. As the number of repetitions of the run interval increases from 3 to 20, the obtained mean accuracy increases from 59.18\% to 66.29\%. This also indicates a positive effect of the run interval on the effectiveness of \compareThree. However, the accuracy of testing results is still no more than 70\%. In terms of reliability, for example, \compareThree enables only 26.92\% of the serverless functions to get the reliable 75\textit{th} percentile performance when the number of repetitions of the run interval is set to 20. As a result, 73.08\% of the functions still cannot get this reliable performance.

The slight increase in the metric values for \compareOne, \compareTwo, and \compareThree may also be due to the fact that when the number of repetitions in the run interval increases, the number of data points in the performance distribution for the initial comparison also increases.
Overall, the effectiveness of baselines are sensitive to the number of repetitions of the run interval, with improved accuracy and reliability as the number increases. 
However, even if the number of repetitions of the run interval is set to 20, the maximum accuracy obtained by \compareOne, \compareTwo, and \compareThree is 75.99\%, 77.15\%, and 66.29\%, respectively. These results cannot reach 80\%.
Moreover, over 60\% of the functions still cannot get reliable percentile performance.


\finding{
Using a strict error constraint as the stopping criterion can improve the accuracy of \toolName.
Even when \compareOne, \compareTwo, and \compareThree apply the strictest constraints, the results obtained are inferior to those of \toolName.
\toolName is insensitive to the number of repetitions of the run interval. As this number increases from 3 to 20, \toolName can provide testing results with accuracy ranging from 96.96\% to 97.53\%, exhibiting a trivial difference of about 0.50\%.
\compareOne, \compareTwo, and \compareThree exhibit sensitivity to the number of repetitions of the run interval. However, even with a large number, the maximum accuracy obtained by \compareOne, \compareTwo, and \compareThree cannot reach 80\%. Moreover, over 60\% of the tested serverless functions cannot get reliable percentile performance.
}

\subsection{RQ3: Flexibility of \toolName}

\begin{figure*}[t]
	\centering
 \includegraphics[width=0.95\textwidth]{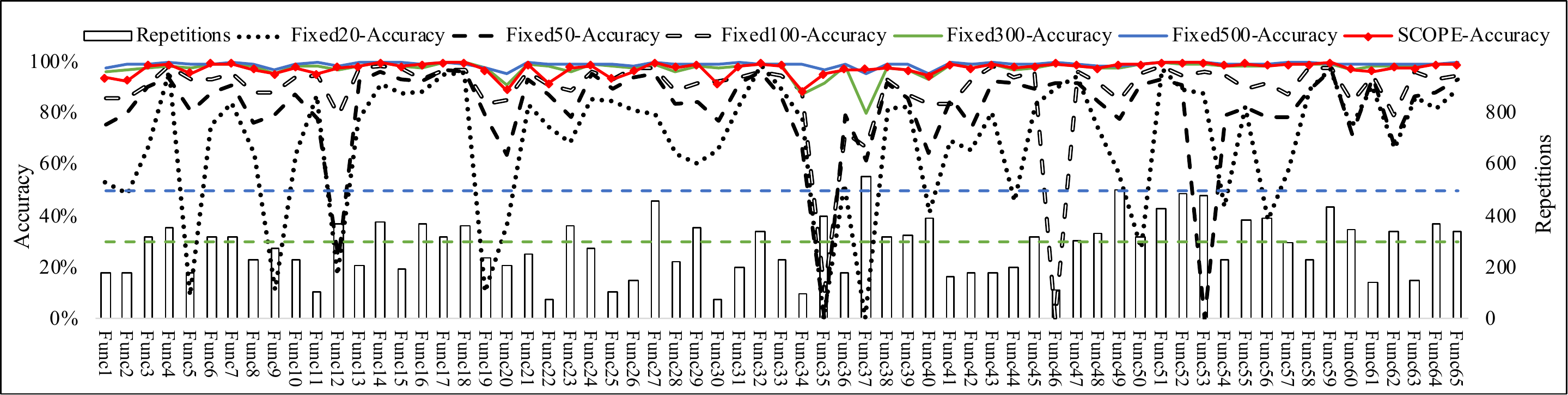}
    \caption{The comparison of the strategy of blindly setting a uniform number of repetitions and \toolName in each tested serverless function (cold starts).}
    \label{fig:compareFixed}
\end{figure*}

We compare \toolName with the strategy of indiscriminately setting a fixed number of repetitions for all evaluated serverless functions. The results show that \toolName is more flexible and effective than this strategy in determining the accurate performance and providing repetitions across various serverless functions.

Fig.~\ref{fig:compareFixed} shows their comparison results in each tested serverless function under cold starts. 
Multiple polylines represent the accuracy results obtained on each serverless function when using the strategy of indiscriminately setting a fixed number of repetitions and \toolName. 
Specifically, when the fixed number of repetitions is blindly set to 20, 50, and 100, the obtained accuracy for more than 95\% of serverless functions is not as high as the accuracy obtained by \toolName for these serverless functions. 
When this fixed number is set to 300, the accuracy results (green polyline) are comparable to those of \toolName.
When this fixed number is set to 500, the accuracy results (blue polyline) may be higher than those of \toolName for most serverless functions.

However, the requirement for repetitions to achieve accurate performance can vary significantly among serverless functions.
The bars in the figure represent the stop location of each serverless function determined by \toolName. 
Some functions may achieve accurate performance with fewer than 100, while others might need over 300 repetitions. It indicates that each serverless function requires a customized repetition to obtain accurate performance. 
It is impractical to rely on a fixed number of repetitions (e.g., 300 or 500) for all evaluated functions. For example, serverless functions that do not require 300 repetitions may require fewer repetitions to obtain accurate performance, then incurring extra running overhead (e.g., 225 additional runs for \textit{Func22}). 
Some functions may require more than 300 repetitions, which makes the results less accurate (e.g., 79.96\% accuracy at 300 repetitions for \textit{Func37}).
For 500-fixed repetitions, although the accuracy results may be higher than those of \toolName on many functions, most serverless functions do not require 500 repetitions to obtain accurate performance, thus incurring significant running overhead.
We measure the overhead by evaluating the total number of repetitions used, which directly correlates with overhead. 
In Fig.~\ref{fig:compareFixed}, the strategy of 500 fixed repetitions necessitates 500*65 = 32,500 repetitions. \toolName requires 18,345 repetitions. Thus, the 500-fixed repetition strategy uses 77.16\% more repetitions of the running overhead than \toolName, which makes \toolName become a cost-efficient option.

\finding{Contrasted with the strategy of indiscriminately setting a fixed number of repetitions, \toolName is more flexible and effective in determining specific repetitions and obtaining accurate performance across various serverless functions.
}


\subsection{Effectiveness to other situations}\label{sec:othersituations}

\begin{table}[t]
\footnotesize
    \caption{The results of other situations using \toolName.}
    \label{tab:other}
    \begin{tabular}{l|p{2cm}|p{1cm}|p{1cm}|p{1cm}|p{1cm}}
        \hline
         \begin{tabular}[c]{@{}l@{}} \end{tabular}  &  \tabincell{c}{\textbf{Accuracy}} & \multicolumn{4}{c} { \tabincell{c}{\textbf{Reliability at 25\textit{th}/50\textit{th}/75\textit{th}/90\textit{th} percentile}}} 
         \\ \hline
         
        \tabincell{c}{Func4  - a mix of cold and warm starts} & 99.47\%	& \checkmark	& \checkmark	& \checkmark &	\checkmark \\

         \hline 
        \tabincell{c}{Func10  - a mix of cold and warm starts} & 95.85\% &	\checkmark	& \checkmark	& \checkmark &	\checkmark \\ \hline 
        \hline 
        \tabincell{c}{App1 from AWS - cold start} & 99.43\% &	\checkmark &	\checkmark & \checkmark	 &	 \checkmark  \\ \hline 
        \tabincell{c}{App1 from AWS - warm start} & 98.82\% &	\checkmark &	\checkmark & \checkmark	 &	 \checkmark  \\ \hline 
        \tabincell{c}{App2 from Google - cold start} & 98.55\% &	\checkmark &	\checkmark & \checkmark	 &	$\times$  \\ \hline 
        \tabincell{c}{App2 from Google - warm start} & 99.77\% &	\checkmark &	\checkmark & \checkmark	 &	\checkmark  \\ \hline 
        \hline 
        \tabincell{c}{Func6 - multiple inputs} & 96.37\% &	\checkmark &	\checkmark &	\checkmark &	 \checkmark  \\ \hline 
        \tabincell{c}{Func7 - multiple inputs} & 97.02\% & \checkmark	 &	\checkmark & \checkmark	 &	 \checkmark  \\ \hline
        \hline
        Func5 - file upload & 98.98\%	& \checkmark &	\checkmark &	\checkmark &	\checkmark \\ \hline 
        Func34 - message queue & 96.04\%	& \checkmark &	\checkmark &	\checkmark &	$\times$\\ \hline
        Func61 - database insert & 97.63\%	& \checkmark &	\checkmark &	\checkmark &	\checkmark \\ \hline
        \hline
        \tabincell{c}{Func32 - bursty workloads} & 99.63\% & \checkmark	 &	\checkmark & \checkmark	 &	 \checkmark  \\ \hline
        \tabincell{c}{Func52 - bursty workloads} & 99.55\% & \checkmark	 &	\checkmark & \checkmark	 &	 \checkmark  \\ \hline
         \hline
        NewFunc1 - Azure - cold start  & 93.61\%	& \checkmark & \checkmark &	\checkmark &	\checkmark \\ \hline 
        NewFunc2 - Azure - cold start & 99.68\%	& \checkmark &	\checkmark &	\checkmark &	\checkmark \\ \hline 
        NewFunc3 - Alibaba - cold start & 97.32\%	& \checkmark &	\checkmark &	\checkmark &	$\times$ \\ \hline 
        NewFunc4 - Alibaba - cold start & 96.68\%	& \checkmark &	\checkmark &	$\times$ &	\checkmark \\ \hline 
        \textbf{Cold start (Mean)} & 96.82\%	& 4/4 &	4/4 &	3/4 &	3/4 \\ \hline 
        \hline
       NewFunc1 - Azure - warm start  & 93.48\%	& \checkmark &	\checkmark &	\checkmark &	\checkmark \\ \hline 
        NewFunc2 - Azure - warm start & 97.01\%	& \checkmark &	\checkmark &	\checkmark &	\checkmark \\ \hline 
        NewFunc3 - Alibaba - warm start  & 95.31\%	& \checkmark &	\checkmark &	\checkmark &	\checkmark \\ \hline 
        NewFunc4 - Alibaba - warm start & 97.09\%	& \checkmark &	\checkmark &	\checkmark &	\checkmark \\ \hline 
        \textbf{Warm start (Mean)} & 95.72\%	& 4/4 &	4/4 &	4/4 &	4/4 \\ \hline 
        
    \end{tabular}
\end{table}

In this section, we explore the effectiveness of \toolName in other situations, including mixed cold and warm conditions, serverless applications composed of multiple functions, varying input conditions, functions with other types of triggers, highly bursty workloads, and functions executed across different platforms. We also explore the effectiveness of \toolName when incorporating additional critical performance metrics, such as tail latency and outlier behavior. 

We explore the effectiveness of \toolName on performance data from serverless functions under mixed cold and warm start conditions. For this purpose, we select two serverless functions: \textit{Func4} and \textit{Func10}. 
Each function is invoked, with the next invocation occurring after a randomly determined time, producing a mix of cold and warm starts, as typically observed in production applications. To establish the ground truth, we collect performance data for each function under 1,000 random executions, thereby capturing a realistic blend of cloud and warm start behavior. Table~\ref{tab:other} shows the results. Specifically, \toolName achieves an accuracy of 99.47\% for \textit{Func4} and 95.85\% for \textit{Func10}, showing high accuracy. Further, \toolName provides reliable 25\textit{th}, 50\textit{th}, 75\textit{th}, and 90\textit{th} percentile performance for these functions. These results demonstrate the effectiveness of our approach in handling mixed cold and warm start performance data for serverless functions.

We explore the effectiveness of \toolName in analyzing serverless applications composed of multiple functions that interact with each other. For this purpose, we implement two serverless applications based on examples provided in the platforms' official documentation. The first serverless application, from the AWS platform~\cite{AWSexamle}, consists of five interacting functions, while the second application, from the Google platform~\cite{Googleexample}, is composed of three functions. These applications are numbered as \textit{App1} and \textit{App2}. We analyze their cold-start and warm-start response latencies using \toolName. To evaluate the effectiveness of \toolName, we collect cold-start and warm-start performance data for each application under 1,000 executions, which serves as the ground truth. Table~\ref{tab:other} shows their results. For cold starts, \toolName achieves an accuracy of 99.43\% for \textit{App1} and 98.55\% for \textit{App2}. Furthermore, \toolName reliably estimates performance at 25\textit{th}, 50\textit{th}, and 75\textit{th} percentile performance for both applications. For warm starts, \toolName achieves an accuracy of 98.82\% for \textit{App1} and 99.77\% for \textit{App2}, demonstrating reliable performance at the 25\textit{th}, 50\textit{th}, 75\textit{th}, and 90\textit{th} percentile percentiles. This illustrates that \toolName can achieve similar effectiveness for serverless application performance.

We explore the effectiveness of \toolName under varying input conditions for serverless functions. For this purpose, we select two serverless functions: \textit{Func6} and \textit{Func7}, which allow for flexible adjustment of input to alter the computation scale. We generate three distinct inputs for each function to ensure a broader evaluation, rather than relying on a single input. Although three inputs provide a reasonable balance between diversity and experimental feasibility, our approach is not limited to this number and can be extended to more inputs if needed.
For example, \textit{Func7} solves linear equations with input $n$ representing the matrix size. We generate three sizes, e.g., 15, 18, and 20. During each warm-start invocation, one of these inputs is randomly selected to produce the corresponding performance data. To evaluate \toolName, we collect performance data for each function over 1,000 executions with these varying inputs, which serves as the ground truth. Table~\ref{tab:other} shows their results. \toolName can achieve an accuracy of 96.37\% for \textit{Func6} and 97.02\% for \textit{Func7}. Moreover, \toolName reliably estimates performance at the 25\textit{th}, 50\textit{th}, 75\textit{th}, and 90\textit{th} percentiles for both functions, demonstrating its effectiveness in handling diverse input conditions.

We explore the effectiveness of \toolName under serverless functions with other types of triggers. For this purpose, we select three functions (\textit{Func5}, \textit{Func34}, and \textit{Func61}) from our set of 65 serverless functions. Then, we modify them to use file uploads, message queues, and database inserts as their triggers by incorporating Amazon S3, Amazon SQS, and Amazon DynamoDB, respectively. We analyze their warm-start response latencies using \toolName. To obtain the end-to-end response time, we calculate the time difference between the initiation of the triggering event and the completion of the function execution. This measurement is consistently performed by recording the event's start time and the function's end time within the function code and then logging the computed time difference. The resulting time difference, referred to as the end-to-end response time, includes both the cloud infrastructure overhead and the function execution time.
We collect performance data for each function across 1,000 executions, which serves as the ground truth. The results are presented in Table~\ref{tab:other}. \toolName achieves an accuracy ranging from 96.04\% to 98.98\%. Moreover, \toolName reliably estimates the performance at 25\textit{th}, 50\textit{th}, 75\textit{th}, and 90\textit{th} percentiles for \textit{Func5} and \textit{Func61}. Furthermore, \toolName provides reliable 25\textit{th}, 50\textit{th}, and 75\textit{th} percentile performance for \textit{Func34}. These results highlight the effectiveness of \toolName in assessing the performance of serverless functions triggered by various event types.

We investigate the effectiveness of \toolName under highly bursty workloads for serverless functions. To this end, we select two serverless functions: \textit{Func32} from AWS Lambda and \textit{Func52} from Google Cloud Functions. These functions are subjected to a series of random burst invocations after prolonged periods of inactivity. The bursty invocations consist of randomly generated concurrent requests. We respectively collect 1,000 performance data triggered by these bursty invocations for both \textit{Func32} and \textit{Func52}, which serve as the ground truth for evaluating the accuracy and reliability of the results produced by \toolName. As shown in Table~\ref{tab:other}, \toolName achieves an accuracy of 99.63\% for \textit{Func32} and 99.55\% for \textit{Func52}. Additionally, \toolName provides reliable performance across the 25\textit{th}, 50\textit{th}, 75\textit{th}, and 90\textit{th} percentiles for both functions. This illustrates that \toolName effectively handles serverless function performance with highly bursty workloads.

We investigate the effectiveness of \toolName on performance data from serverless functions executed on different platforms. Specifically, we conduct experiments on serverless functions deployed on Microsoft Azure Functions and Alibaba Function Compute. Four serverless functions are implemented: two from Microsoft Azure Functions (denoted as \textit{NewFunc1} and \textit{NewFunc2}) and two from Alibaba Function Compute (denoted as \textit{NewFunc3} and \textit{NewFunc4}). We analyze their cold-start and warm-start response latencies using \toolName. To evaluate the effectiveness of \toolName, we collect cold-start and warm-start performance data for each function over 1,000 executions, which serve as the ground truth for comparison. The results are presented in Table~\ref{tab:other}. For cold starts, \toolName achieves a mean accuracy of 96.82\% across these functions. Additionally, \toolName reliably captures performance at the 25\textit{th}, 50\textit{th}, 75\textit{th}, and 90\textit{th} percentiles for four, four, three, and three functions, respectively. For warm starts, \toolName achieves a mean accuracy of 95.72\%, with reliable performance at the 25\textit{th}, 50\textit{th}, 75\textit{th}, and 90\textit{th} percentiles for all four functions, respectively. These results demonstrate the effectiveness and generalizability of \toolName in assessing the performance of serverless functions executed on different serverless platforms.

\begin{table}[t]
\footnotesize
    \caption{The results of \toolName considering other performance metrics.}
    \label{tab:scopeNew}
    \begin{tabular}{l|p{1.5cm}|p{2cm}|p{2cm}|p{2cm}|p{2cm}}
        \hline
         \begin{tabular}[c]{@{}l@{}} \end{tabular}  &  \tabincell{c}{\textbf{Accuracy}} & \multicolumn{4}{c} { \tabincell{c}{\textbf{\textit{\#Functions}: reliability at 25\textit{th}/50\textit{th}/75\textit{th}/90\textit{th} percentile}}} 
         \\ \hline
        \textbf{\textit{\toolName}} & 97.25\% &	87.69\%	& 93.08\% &	92.31\%	& 90.77\% \\ 
        \hline
        \textbf{\textit{\toolName}} + Tail & 98.84\% &	91.54\% &	96.92\%	& 94.62\% &	93.08\% \\ 
        \hline
        \textbf{\textit{\toolName}} + Outlier & 97.61\% &	88.46\%	& 93.08\% &	92.31\%	& 90.00\% \\ 
        \hline
    \end{tabular}
\end{table}

We explore the effectiveness of \toolName when incorporating additional critical performance metrics, such as tail latency and outlier behavior. Specifically, we consider the tail latency of the 95\textit{th} percentile performance in \toolName and check whether its confidence interval falls within a defined error margin (e.g., 3\%) of the observed true percentile performance. Since enforcing constraints on tail latency is even more difficult to achieve, the error threshold may need to be relaxed compared to the original constraint settings for non-tail latency. This is because tail latency is inherently more variable and sensitive, making strict enforcement more challenging in practice.
Additionally, we investigate the inclusion of outlier behavior in \toolName. Outliers are determined using the Interquartile Range (IQR) method, a statistical approach for detecting outliers. In this method, outliers are identified as data points that fall outside the range defined by the first and third quartiles, typically beyond 1.5 times the IQR. If the number of outliers does not exceed a predefined threshold (e.g., 10\% of the total test data), the outlier behavior is considered negligible in our study. Table~\ref{tab:scopeNew} shows the mean results of two new versions of \toolName in cold-start and warm-start performance testing for serverless functions. The mean accuracy achieved by \toolName considering tail latency is 98.84\%, and 97.61\% for outlier behavior. Regarding reliability, \toolName with tail latency provides reliable 25\textit{th}, 50\textit{th}, 75\textit{th}, and 90\textit{th} percentile performance for 91.54\%, 96.92\%, 94.62\%, and 93.08\% of the serverless functions, respectively. Similarly, \toolName with outlier behavior provides reliable 25\textit{th}, 50\textit{th}, 75\textit{th}, and 90\textit{th} percentile performance for 88.46\%, 93.08\%, 92.31\%, and 90.00\% of the functions, respectively. From these results, adding other critical performance metrics, such as tail latency or outlier behavior, can improve the effectiveness of \toolName. It is important to note that even without the integration of these critical metrics, our approach maintains consistency and high effectiveness. This suggests that while adding tail latency and outlier behavior improves the robustness and comprehensiveness of the analysis, \toolName already performs well across different serverless functions.

\section{Discussion}

\subsection{Why \toolName?}
Similar to \compareOne, \compareTwo, and \compareThree, \toolName efficiently determines the number of repetitions for each serverless function, ensuring accurate and reliable performance testing outcomes. Notably, \toolName achieves an impressive accuracy rate of 97.25\%, outperforming \compareOne, \compareTwo, and \compareThree by 44.97, 33.83, and 36.83 percentage points, respectively. The ineffectiveness of baselines stems from fundamental differences in approach, which fail to accommodate the unique characteristics of serverless functions.

The question then arises: \emph{does \toolName's success solely depend on mandating a higher number of repetitions?} The answer is no. While heightened accuracy in performance testing inherently demands increased repetitions, \toolName refines this approach by ensuring the effective determination of additional repetitions. This is evidenced by our analysis in RQ3. Specifically, in RQ3, we adopt a fixed number of 300 repetitions per serverless function, which maintains a comparable total volume of repetitions to \toolName across all functions considered and achieves a similar overall accuracy. However, this comparison exposes the inefficiencies inherent in employing a fixed number. Employing a rigidly high fixed repetition number for serverless functions that do not require 300 repetitions produces unnecessary resource usage and prolonged testing times, without a commensurate increase in result precision. Conversely, this fixed approach proves inadequate for functions that may require more repetitions, risking insufficient coverage and potentially failing to guarantee the accuracy and reliability of serverless function performance.


Therefore, while high-accuracy performance testing inherently demands a greater number of repetitions, \toolName surpasses this requirement by adopting a strategy that does not just increase repetitions but does so with discerning efficiency. This strategy ensures \toolName's efficacy across a diverse array of serverless functions, validating its effectiveness. Thus, \toolName's superior effectiveness is not merely a product of increased testing iterations but results from a strategic and judicious application of those additional repetitions, tailored to the unique demands of each serverless function it evaluates.

\subsection{Validity of ground truth performance}\label{sec:groundtruth}

We use 1,000 repetitions as the ground truth performance for each serverless function to identify the technique's effectiveness. This may raise the question of whether a sufficiently large number of repetitions are chosen for ground truth performance. To address it, we execute each serverless function for an additional 500 runs. When using execution results of 1,000, 1,100, 1,200, 1,300, 1,400, and 1,500 repetitions of the ground truth, the mean accuracy of \toolName is between 96.18\% and 97.25\%, a difference of about one percentage point. The consistent results indicate that using 1,000 executions has established a trustworthy ground truth. An additional 500 runs are added to our online repository~\cite{ourdata}.

\subsection{Discussion of different design, data, and parameter choices}

\toolName is built with a flexible, effective, and adaptable design aimed at providing accurate performance analysis across a wide range of serverless functions. Central to the design is the stopping criterion, which evaluates the tested performance data of serverless functions. This criterion is based on non-parametric confidence interval (CI) calculations, and we support three mainstream CI calculation methods to demonstrate the flexibility, applicability, and effectiveness of our approach. Furthermore, as discussed in Section~\ref{sec:othersituations}, we explore the inclusion of additional critical performance metrics, such as tail latency and outlier behavior, into the design of \toolName. The results demonstrate that while incorporating these metrics improves the results of the analysis, \toolName performs well across different serverless functions, achieving an accuracy of 97.25\%. This indicates that even without these additional metrics, \toolName remains highly effective for performance testing. 

\toolName is evaluated using a diverse dataset that includes 25 distinct task types. These tasks cover a broad array of applications, including mathematical operations, image processing, face detection, graph network analysis, video processing, and natural language processing. This diverse set of functions ensures that our evaluation reflects a wide range of real-world serverless computing workloads, enabling us to assess the generalizability of \toolName across different domains and use cases, both in cold-start and warm-start scenarios. The additional situations explored in Section~\ref{sec:othersituations} further confirm its applicability across various situations, such as mixed cold and warm conditions, serverless applications composed of multiple functions, varying input conditions, functions with other types of triggers, highly bursty workloads, and functions executed across different platforms. Additionally, we use a trustworthy ground truth to validate the performance of \toolName. To assess the sufficiency of 1,000 repetitions as the ground truth, we conduct an additional 500 runs for each function, as discussed in Section~\ref{sec:groundtruth}. The results show that the effectiveness of \toolName remains consistent when compared to the 1,500 repetitions.

\toolName's evaluation also includes a thorough investigation of parameter choices. In RQ2, we examine the impact of varying key parameters, such as different constraints for the stopping criterion and varying the number of repetitions within each run interval. The results demonstrate that \toolName continues to maintain high effectiveness regardless of these parameter variations. Additionally, we assess \toolName by comparing how well the estimated performance aligns with the true performance across different percentiles (25\textit{th}, 50\textit{th}, 75\textit{th}, 90\textit{th}). This analysis highlights the ability of \toolName.

In summary, the design, data, and parameter choices of \toolName are carefully tailored to provide an effective performance testing approach across a variety of serverless functions.

\subsection{Discussion of implications}

Our work provides several key implications for the design and implementation of performance testing approaches, spanning the pre-design, design, and post-design phases. These implications are outlined as follows:
(1) When designing a performance testing approach, it is essential to assess the magnitude of the performance data in order to determine the most appropriate level of granularity for analysis. If the magnitude is large, a coarse-grained analysis might be more efficient, as it provides a broader overview with less computational overhead. Conversely, for smaller magnitude, a fine-grained analysis may be necessary to capture subtle variations in performance. This consideration helps to target the accuracy in the performance evaluation process. (2) Another critical consideration is the distribution of the performance data. Different analysis methods are suitable depending on whether the data follows a normal or non-normal distribution. In cases where performance data exhibits a non-normal distribution, non-parametric methods are the most appropriate choice. These methods do not assume any specific data distribution, making them more robust for handling irregular data patterns. This underscores the need for careful method selection to ensure reliable and valid performance analysis. (3) Performance characteristics can vary significantly depending on the task type being executed. Therefore, any performance testing approach must incorporate a diverse set of task types to ensure a comprehensive evaluation. By testing across a wide range of tasks, one can ensure that the approach remains effective in a variety of real-world tasks. This variability in performance highlights the importance of adapting the performance testing approach to different contexts and workloads to ensure generalizability.

\subsection{Threats to validity}\label{sec:discussion}

\noindent \textbf{Selection of baseline methods.}  
Embracing serverless computing enables developers to create cloud applications in a new programming paradigm. However, the literature on serverless computing lacks dedicated performance testing approaches. As a result, baselines specifically designed for serverless performance testing are not currently available. To address this, we evaluate serverless function performance using performance testing methods designed for traditional cloud applications. This may raise a threat to the representativeness of baseline methods. To mitigate this threat, we select three state-of-the-art performance testing methods: \compareOne, \compareTwo, and \compareThree. They have demonstrated superiority over previous performance testing techniques for cloud applications~\cite{he2019statistics, he2021performance}. These methods are detailed in Section~\ref{sec:baselines} and Section~\ref{sec:relatedwork}. Therefore, our selected baseline methods are representative of the state-of-the-art.

\noindent \textbf{Conclusion of technical effectiveness.}
We evaluate the effectiveness of  \toolName and state-of-the-art techniques on the performance results of serverless functions. 
Technical effectiveness may vary depending on the types of serverless functions. This may potentially influence the experimental conclusions that we summarize herein. 
To mitigate it, we investigate the performance of 65 serverless functions from a publicly available dataset curated in previous work~\cite{wen2023revisiting}, which covers a variety of tasks, e.g., video processing, machine learning, and natural language processing. Thus, the conclusion of technical effectiveness is based on testing results for various types of serverless functions. Although we cannot generalize our results to all serverless functions, the ones used herein are representative of those widely used in previous work.



\subsection{Limitations of \toolName}


While \toolName provides accurate and effective performance evaluations, it currently does not delve into specific distribution characteristics, such as identifying patterns, which could further enhance the depth of performance assessment.
\toolName employs a non-parametric approach for evaluating serverless function performance. This method is advantageous as it makes minimal assumptions about the underlying probability distribution of performance data, making it well-suited for scenarios where the distribution is unknown or highly variable. However, if the performance distribution of serverless functions is explicitly determined and incorporated into the analysis, it could enable \toolName to provide even more nuanced insights and potentially improve its evaluation effectiveness.



\subsection{Challenges of \toolName}

In the process of designing \toolName, we face several key challenges related to selecting appropriate checks, statistical methods, and performance metrics for analysis. 
First, in serverless computing, it is crucial to determine the appropriate level of granularity for performance analysis. Due to the latency characteristics of serverless functions, a fine-grained approach is necessary to capture subtle changes in performance data and ensure effective analysis. Second, serverless functions generally exhibit irregular performance patterns. This variability necessitates careful consideration of which statistical methods are most appropriate. Since the performance data of many serverless functions, both in cold and warm starts, follows a non-normal distribution, non-parametric methods are more suitable for accurate performance analysis because they require fewer distribution assumptions. Finally, once the appropriate statistical methods are identified, we must determine which performance metrics should be analyzed. A critical aspect of our approach is assessing whether the performance metric falls within a desired confidence interval. Given the fine-grained nature of our analysis, we extend our checks to cover a broader range of the performance distribution—specifically from the 25\textit{th} percentile to the 75\textit{th} percentile. This ensures that results are reliable and accurate.

\subsection{Future work}



In future work, we aim to extend the capabilities of \toolName by testing a wider variety of serverless functions and applications across different serverless providers. This will involve conducting performance evaluations in diverse real-world scenarios, such as hybrid cold-warm start environments, variations in input types, and different event triggers. Additionally, we plan to integrate \toolName with various monitoring and profiling tools to gain deeper insights into the runtime behavior of serverless functions, enhancing the diagnostic and optimization capabilities of our approach.

Furthermore, we intend to develop \toolName into a serverless API, making it easily accessible for developers and researchers to perform performance evaluations of serverless functions and applications. The API will provide customizable testing parameters, allowing for flexible performance analysis in various contexts. By providing seamless access to \toolName, we hope to promote its wider adoption and support the continuous evolution of performance testing techniques within the serverless computing domain.


\section{Related Work}\label{sec:relatedwork}

\noindent\textbf{Performance of serverless computing.} Performance is the most studied topic in the serverless computing literature~\cite{wen2022literature, li2022serverless, scheuner2020function}. On one hand, researchers have proposed novel solutions for optimizing the performance of serverless functions~\cite{wen2023FaaSlight, OakesATC18-140, mahgoub2021sonic-128, lin2020modeling, singhvi2021atoll, li2023dataflower, mcgrath2017serverless, qi2023halfmoon, jin2023ditto}.
For example, \textit{FaaSLight}~\cite{wen2023FaaSlight} loaded only indispensable code for serverless functions to improve overall performance. \textit{SOCK}~\cite{OakesATC18-140} cached commonly used libraries to speed up the cold-start performance of serverless functions. A management layer called \textit{SONIC}~\cite{mahgoub2021sonic-128} was designed to improve the communication performance between serverless functions. 
On the other hand, empirical studies have delved into characterizing the performance of serverless computing. For instance, Eismann \textit{et al.}~\cite{eismann2022case} utilized a case study on AWS Lambda to investigate the stability of performance measurements by exploring the impact of various configuration settings. 
McGrath\textit{et al.}~\cite{mcgrath2017serverless} designed a performance-oriented serverless platform and evaluated the performance characteristics of serverless platforms.
Wen~\textit{et al.}~\cite{wen2021characterizing} conducted a thorough measurement study to characterize the performance of serverless functions executed on different commodity serverless platforms.
However, there is no performance testing approach specifically tailored to serverless computing to help researchers and engineers determine accurate and reliable performance for serverless functions. To fill the gap, we propose \toolName.

\noindent\textbf{Performance testing of cloud applications.}
Researchers have proposed the related performance testing work for traditional cloud applications~\cite{maricq2018taming, wang2018testing, he2019statistics, he2021performance, laaber2019software, liu2023legodroid}. 
Laaber \textit{et al.}~\cite{laaber2019software} explored the impact of cloud environments on the variability of performance test results and assessed the reliability of detecting slowdowns.
Eismann \textit{et al.}~\cite{eismann2020microservices} primarily conducted a measurement study of microservices’ performance and discussed related challenges without proposing a specific performance testing approach to determine the accurate performance.
Wang \textit{et al.}~\cite{wang2018testing} proposed a non-parametric approach for cloud performance testing, based on basic bootstrapping techniques. Similarly, Maricq \textit{et al.}~\cite{maricq2018taming} introduced CONFIRM, a non-parametric method that employed bootstrapping for cloud performance testing. However, He \textit{et al.}~\cite{he2021performance} demonstrated that the bootstrapping method lacks the consideration of internal data dependency, causing a low accuracy in cloud application performance. To address this limitation, He \textit{et al.}~\cite{he2021performance} improved this kind of approach by incorporating the block bootstrapping method, which accounted for the internal data dependency.

For the stopping criterion of performance testing, Alghmadi \textit{et al.}~\cite{alghmadi2016automated} measured the degree of repetition of data in performance results. 
However, previous work~\cite{he2019statistics} showed that this stopping criterion was not appropriate for performance testing of cloud applications. 
\compareOne~\cite{he2019statistics} and \compareTwo~\cite{he2021performance} have considered the stability of performance distributions to terminate the repeated runs, and have been evaluated to be the state-of-the-art in performance testing for cloud applications~\cite{he2021performance, eismann2020microservices}.
However, our evaluation uncovers that \compareOne and \compareTwo show low effectiveness in serverless computing. This paper proposes \toolName with a novel stopping criterion, which considers accuracy and consistency checks and outperforms \compareOne and \compareTwo in the performance testing for serverless functions.

\section{Conclusion}\label{sec:conclusion}
This paper explores performance testing in the serverless computing domain. We proposed \toolName, the first performance testing approach specifically tailored for serverless computing, which included accuracy and consistency checks. \toolName emphasized the need for accuracy for most performance of serverless functions to determine accurate and reliable performance.
We investigated 65 serverless functions and used their performance results to evaluate the effectiveness of \toolName and state-of-the-art techniques.
The results showed that \toolName provided testing results with 97.25\% accuracy, 33.83 percentage points higher than the best currently available technique.

\begin{acks}
This work is supported by the National Natural Science Foundation of China under Grant Nos. 62325201, 62425203, and 62032003, Beijing Natural Science Foundation under Grant No. 4244076, and Beijing Postdoctoral Research Foundation under Grant No. 2024-ZZ-20.
\end{acks}

\bibliographystyle{ACM-Reference-Format}
\bibliography{main}

\end{document}